\documentclass[twocolumn,superscriptaddress,aps,jcp,amsmath,amssymb]{revtex4-1}

\usepackage[latin9]{inputenc}
\usepackage{amsthm}
\usepackage{amsmath}
\usepackage{amssymb}
\usepackage{graphicx}
\usepackage{esint}
\usepackage{hyperref}
\usepackage{paralist}
\usepackage{xcolor} 
\usepackage{epstopdf}

\graphicspath{{Figures/}}

\def\C{{\mathbb C}}
\def\R{{\mathbb R}}
\def\RR{{\mathbb{R}}}
\def\CC{{\mathbb{C}}}

\def\ge{\geqslant}

\newcommand{\wh}{\widehat}
\newcommand{\wt}{\widetilde}
\newcommand{\ud}{\,\mathrm{d}}

\newcommand{\Or}{\mathcal{O}}
\newcommand{\bd}{\boldsymbol}

\theoremstyle{plain}

\theoremstyle{definition}

\newtheorem*{remark*}{Remark}

\newcommand{\mc}[1]{\mathcal{#1}}

\newcommand{\abs}[1]{\lvert#1\rvert}
\newcommand{\Abs}[1]{\left\lvert#1\right\rvert}

\newcommand{\norm}[1]{\lVert#1\rVert}

\newcommand{\bra}[1]{\langle#1\rvert}

\newcommand{\ket}[1]{\lvert#1\rangle}

\newcommand{\dps}{\displaystyle}

\DeclareMathOperator{\tr}{Tr}

\begin{document}

\title{Path integral molecular dynamics with surface hopping for thermal equilibrium sampling of nonadiabatic systems}

\author{Jianfeng Lu} \email{jianfeng@math.duke.edu}
\affiliation{Department of Mathematics, Duke University, Box 90320,
Durham NC 27708, USA}\affiliation{Department of Physics and Department
of Chemistry, Duke University, Durham NC 27708, USA} \author{Zhennan
Zhou} \affiliation{Department of Mathematics, Duke University, Box
90320, Durham NC 27708, USA}

\begin{abstract} In this work, a novel ring polymer representation for
  multi-level quantum system is proposed for thermal average
  calculations. The proposed representation keeps the discreteness of
  the electronic states: besides position and momentum, each bead in
  the ring polymer is also characterized by a surface index indicating
  the electronic energy surface. A path integral molecular dynamics
  with surface hopping (PIMD-SH) dynamics is also developed to sample
  the equilibrium distribution of ring polymer configurational space.
  The PIMD-SH sampling method is validated theoretically and by
  numerical examples.
\end{abstract}

\maketitle

\section{Introduction}

The development of efficient methods to simulate complex chemical
systems on the quantum level has been a central challenge in
theoretical and computational chemistry. Exact quantum simulation of
coupled nuclear and electronic system is numerically formidable even
for small molecules, and, therefore, approximate methods are needed
with a reasonable balance of computational effort and incorporating
some quantum mechanical aspects of the dynamics.

Under the classical Born-Oppenheimer approximation, one can separate
the degrees of freedom associated to the nuclei and electrons, so that
a Hamiltonian operator consists of kinetic term and a potential energy
surface can be obtained for the nuclear degrees of freedoms. However,
when the nonadiabatic effect can not be neglected (often referred as
the regime of beyond Born-Oppenheimer dynamics), we need to explicitly
include multi-level electronic states in the Hamiltonian, and thus
more than one energy surfaces corresponding to different electronic
states have to be incorporated.  We refer the readers to the reviews
\cite{Makri:99, StockThoss:05, Kapral:06} for general discussions on
simulation methods in the nonadiabatic regime.  In this work, we
focus on the thermal averages like $\tr[e^{-\beta \wh{H}} \wh{A}]$,
where $\wh{H}$ is a multi-level Hamiltonian operator, $\wh{A}$ is an
observable, and $\beta$ is the inverse temperature.

For the thermal average calculation, the ring polymer representation,
based on the imaginary time path integral, has been a popular approach
to map a quantum particle in thermal equilibrium to a fictitious
classical ring polymer in copies of the phase space
\cite{Feynman:72}. The representation is asymptotically exact as the
number of beads in the ring polymer goes to infinity. Based on the
ring polymer representation, path integral Monte Carlo (PIMC)
\cite{ChandlerWolynes:81, BerneThirumalai:86} and path integral
molecular dynamics (PIMD) \cite{MarklandManolopoulos:08,
  CeriottiParrinelloMarklandManolopoulos:10} sampling techniques have
been developed to calculate the quantum statistical average. The ring
polymer representation has also been used in the dynamics simulations,
such as, the centroid molecular dynamics \cite{CaoVoth:94a,
  CaoVoth:94b, JangVoth:99}, the ring polymer molecular dynamics
\cite{CraigManolopoulos:04, HabershonManolopoulosMarklandMiller:13},
Matsubara dynamics \cite{HeleWillattMuoloAlthorpe:15a,
  HeleWillattMuoloAlthorpe:15b}, and path integral Liouville dynamics
\cite{Liu2014, LiuZhang2016}.

The conventional ring polymer representation however only works in the
adiabatic regime. To apply methods like path-integral molecular
dynamics (or dynamic extensions like ring-polymer molecular dynamics)
to multi-level systems when the nonadiabatic effects cannot be
neglected, a popular strategy is to use the mapping variable approach
\cite{MeyerMiller:79, StockThoss:97}, see also the review article
\cite{StockThoss:05} and more recent developments in
\cite{AnanthMiller:10, RichardsonThoss:13,
  Ananth:13,MenzeleevBellMiller:14,CottonMiller:15,
  HeleAnanth:16,JianLiu:16}.  The idea is to replace the multi-level
system by a single level system with higher dimension by mapping the
discrete electronic states to continuous variables using uncoupled
harmonic oscillators \cite{StockThoss:97}. The ring polymer
representation can then be applied to the mapped system
\cite{AnanthMiller:10, RichardsonThoss:13,
  Ananth:13,MenzeleevBellMiller:14, HeleAnanth:16}. 

In this work, we consider rather an alternative
  strategy for extending the ring polymer representation to
  multi-level systems, following the spirit of the pioneering work of
  Schmidt and Tully \cite{SchmidtTully2007}. In the proposed
  representation, each bead in the ring polymer is associated with a
  surface index indicating which energy surface the bead lies on. The
  total Hamiltonian, and thus the sampling, is given in the extended
  phase space of position, momentum and surface index of each bead in
  a ring polymer. While \cite{SchmidtTully2007} uses the adiabatic
  picture, the idea can be generalized to other basis, and this work
  uses the diabatic picture (which also recovers the ring polymer
  representation in \cite{SchmidtTully2007} in adiabatic picture).

  As another main contribution of our current work, we propose a path
  integral molecular dynamics with surface hopping method (PIMD-SH) to
  efficiently sample the ring polymer representation, where the
  discrete electronic state is sampled via a consistent surface
  hopping algorithm coupled with Hamiltonian dynamics of the position
  and momentum with Langevin thermostat. It is shown that the PIMD-SH
  method ergodically samples the correct equilibrium distribution on
  the extended phase space and can thus be used to sample multi-level
  quantum systems. In addition, effective numerical integrators are
  proposed for PIMD-SH to simulate the ergodic trajectories. The
  numerical results validate the proposed PIMD-SH method for thermal
  equilibrium sampling of multi-level quantum systems.

Compared with the approaches based on mapping variables, the PIMD-SH
is more direct and treats the discrete electronic variables explicitly
in the sampling. We think this is more advantageous than treating the
electronic and nuclear degrees of freedom on the same footing, as it
allows us to employ numerical strategies that exploit the scale
separation between the nuclei and electrons and also offer more
flexibility. As another advantage, since we use a surface hopping type
dynamics to treat the discrete electronic states, it is more natural
to combine the proposed thermal (imaginary time) sampling method with
real time surface hopping dynamics \cite{Tully:90,
  HammesSchifferTully:94, Tully:98, Kapral:06, ShenviRoyTully:09,
  Barbatti:11, SubotnikShenvi:11, Subotnik:16}, which is one of the
motivations of our development following our recent works in surface
hopping dynamics \cite{FGASH,FGASH2}.  Let us remark that there have
been recent works trying to combine the path integral formulation and
surface hopping dynamics \cite{ShushkovLiTully:12}, though it is
unclear if the trajectory with hopping dynamics can preserve the
thermal equilibrium. See also discussions on equilibrium properties of
the fewest switches surface hopping dynamics in
\cite{SchmidtParandekarTully:08, ShermanCorcelli:15}.

The paper is outlined as follows. In Section \ref{sec:theory}, we
present the ring polymer representations for the thermal averages for
multi-level quantum systems. The detailed derivation of the ring
polymer representation is given together with the proposed PIMD-SH
dynamics to sample the equilibrium distribution on the extended phase
space. The numerical integration of the PIMD-SH dynamics is discussed
in Section \ref{sec:algorithm}, where we combine a surface hopping
dynamics and the Langevin thermostat to treat the electronic states
and phase space variables. The numerical tests are presented in
Section \ref{sec:test} to validate the performance of the PIMD-SH
method. In the conclusion and Appendix, we discuss possible future
directions and in particular the generalization to more general
Hamiltonians.

\section{Theory} \label{sec:theory}

\subsection{The ring polymer representation for canonical distribution
for two-level systems} \label{sec:ringpolymer}

For simplicity of notation, we will restrict the presentation to
two-level systems, while the methodology can be generalized to
multi-level systems, which will be deferred to  Appendix~\ref{App:B}.  In a diabatic representation, the Hamiltonian
operator of a general two-level system can be written as (atomic unit
is used)
\[ \wh H = \wh T+ \wh V = \frac{1}{2M}
\begin{pmatrix} \wh p^2 & \\ & \wh p^2
\end{pmatrix} +
\begin{pmatrix} V_{00}(\wh q) & V_{01}(\wh q) \\ V_{10}(\wh q) &
V_{11} (\wh q)
\end{pmatrix},
\] where $\wh q$ and $\wh p$ are the nuclear position and momentum
operators, and $M$ is the mass of the nuclei (for simplicity we assume
all nuclei have the same mass). At any position $q \in \RR^d$, the
matrix potential
\begin{equation*} V(q) = \begin{pmatrix} V_{00}(q) & V_{01}(q) \\
V_{10}(q) & V_{11}(q)
    \end{pmatrix}
\end{equation*} is a Hermitian matrix which for example comes from the
projection of the electronic Hamiltonian to two low-lying states.
Here for simplicity, we will assume that the off-diagonal potential
functions $V_{01} = V_{10}$ are real valued ($V_{00}$ and $V_{11}$ are
real since $V$ is Hermitian). In addition, we will consider the
simpler case that the off-diagonal term $V_{01}(q)$ does not change
sign for all $q$, the formulation for the general case is discussed in
the Appendix. Let us remark that the same simplifying assumptions are
made explicitly or implicitly also in the mapping variable approaches
(see e.g., \cite{Ananth:13,MenzeleevBellMiller:14}).

The Hilbert space of the system is thus $L^2(\RR^d) \otimes \CC^2$,
where $d$ is the spatial dimension of the nuclei position degree of
freedom. In this work, we consider the thermal equilibrium average of
observables, given by
\begin{equation}\label{eq:aveA} \langle\wh{A}\rangle = \frac{\tr_{ne}
[e^{-\beta \wh H} \wh A ] }{\tr_{ne}[e^{-\beta \wh H}]},
\end{equation} for an operator $\wh A$, where $\beta = \frac{1}{k_B
T}$ with $k_B$ the Boltzmann constant and $T$ the absolute
temperature, and $\tr_{ne}$ denotes trace with respect to both the
nuclear and electronic degrees of freedom, namely,
\[ \tr_{ne}=\tr_{n}\tr_{e}=\tr_{L^2 (\R^d)}\tr_{\C^{2}}.
\] The denominator in \eqref{eq:aveA} is the partition function given
by $\mc{Z} = \tr_{ne}[e^{-\beta \wh H}]$. 
For simplicity, throughout this work, we will
assume that the observable $\wh A$ only depends on $q$, but not $p$,
in other words, $\wh A$ can be written as
\begin{equation*} \wh{A} = \begin{pmatrix} A_{00}(\wh{q}) &
A_{01}(\wh{q}) \\ A_{10}(\wh{q}) & A_{11}(\wh{q}) \end{pmatrix}.
\end{equation*}

\begin{figure}[ht]
\begin{centering}
\includegraphics[scale=0.5]{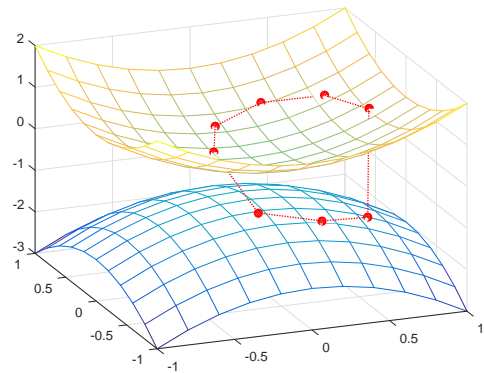}
\caption{Schematic plot of a ring polymer on the extended phase space
with two energy surfaces.}
\label{fig:beads}
\end{centering}
\end{figure}

As we will show in Section~\ref{sec:derivation}, for a sufficiently
large $N \in \mathbb N$, we may approximate the partition function by
a ring polymer representation with $N$ beads
\begin{multline}\label{eq:reZ} \tr_{ne}[e^{-\beta \wh H}] \approx
\mc{Z}_N := \frac{1}{(2\pi)^{dN}} \int_{\RR^{2dN}} \ud \bd q \ud \bd p
\sum_{\bd{\ell} \in \{0, 1\}^N} \\ \times \exp(-\beta_N H_N(\bd{q},
\bd{p}, \bd{\ell})).
\end{multline} where $\beta_N=\beta/N$. The ring polymer that consists of
$N$ beads is prescribed by the configuration $(\bd{q}, \bd{p},
\bd{\ell}) \in \RR^{2dN} \times \{0, 1\}^N$, where $\bd{q} = (q_1,
\cdots, q_N)$ and $\bd{p} = (p_1, \cdots, p_N)$ are the position and
momentum of each bead, and $\bd{\ell} = (\ell_1, \cdots, \ell_N)$
indicates the energy level of the bead (thus each bead in the ring
polymer lives on two copies of the classical phase space $\RR^{2d}$,
see Figure~\ref{fig:beads} for an illustration).  For a given ring
polymer with configuration $(\bd{q}, \bd{p}, \bd{\ell})$, the
effective Hamiltonian $H_N(\bd{q}, \bd{p}, \bd{\ell})$ is given by
\begin{equation}\label{eq:Ham} H_N(\bd q, \bd p, \bd\ell) =
\sum_{k=1}^N \bra{\ell_k} G_k \ket{\ell_{k+1}},
\end{equation} where we take the convention that $\ell_{N+1} = \ell_1$
and matrix elements of $G_k$, $k = 1, \ldots, N$, are given by
\begin{widetext}
\begin{equation} \label{eq:Gele} \bra{\ell} G_{k} \ket{\ell'} =
 \begin{cases} \dps \frac{p^2_k}{2M}+ \frac{M \left(q_k-q_{k+1}
\right)^2}{2(\beta_N)^2} +V_{\ell\ell}(q_k) - \frac{1}{\beta_N} \ln
\Bigl( \cosh \bigl( \beta_N \Abs{ V_{01} (q_k) }\bigr) \Bigr), &
\ell=\ell', \\ \dps \frac{p^2_k}{2M}+ \frac{ M\left(q_k-q_{k+1}
\right)^2 }{2(\beta_N)^2} +\frac{V_{00}(q_k)+V_{11}(q_k)}{2}
-\frac{1}{\beta_N}\ln\Bigl( \sinh \bigl(\beta_N \Abs{V_{01}(q_k)}
\bigr) \Bigr), & \ell \ne \ell'.
 \end{cases}
\end{equation}
\end{widetext} Compared to the conventional ring polymer for a single
potential energy surface, the difference in the two-level case is that
now each bead is associated with a level index $\ell_k$. In
particular, when $\ell_k \ne \ell_{k+1}$, two consecutive $k$-th and
$(k+1)$-th beads in the ring polymer stay on different energy
surfaces.  We will call this a kink in the ring polymer. Note that the
number of kinks is always even since the beads form a ring. In
Figure~\ref{fig:beads}, a schematic plot of a ring polymer of $8$
beads with two kinks is shown, where $5$ beads are on the upper energy
surface and $3$ are on the lower energy surface.
 Moreover, notice that if the off-diagonal terms of the
matrix potential $V_{01} = 0$, the diagonal part of $G_k$ falls back
to the usual term in standard ring polymer representation; the current
representation is thus a natural extension to the multi-level case.

For an observable $\wh{A}$, under the ring polymer representation, we
have
\begin{multline} \label{eq:reA} \tr_{ne}[e^{-\beta \wh H}\wh A ]
\approx \frac{1}{(2\pi)^{dN}} \int_{\RR^{2dN}} \ud \bd q \ud \bd p
\sum_{\bd{l} \in \{0, 1\}^N} \\ \times \exp(-\beta_N H_N) W_N[A],
\end{multline} where the weight function associated to the observable
is given by (recall that $\wh{A}$ only depends on position by our
assumption)
\begin{multline}\label{eq:WNA} W_N[A] (\bd q, \bd p, \bd\ell) =
  \frac{1}{N} \sum_{k=1}^N \biggl( \bra{\ell_k} A(q_k) \ket{\ell_k} - \\
  - e^{\beta_N \langle \ell_k| G_k| \ell_{k+1} \rangle - \beta_N
    \bra{\bar{\ell}_k} G_{k} \ket{ \ell_{k+1}}} \bra{\ell_{k}}
  A(q_{k}) \ket{\bar{\ell}_k}
  \frac{V_{\ell_k\bar{\ell}_k}}{\abs{V_{\ell_k\bar{\ell}_k}}} \biggr),
\end{multline}
where we have introduced the short hand notation
$\bar{\ell}_k = 1 - \ell_k$, \textit{i.e.}, $\bar{\ell}_k$ is the
level index of the other potential energy surface than the one
corresponds to $\ell_k$ in our two-level case.  Similar as for the
partition function, the ring polymer representation \eqref{eq:reA}
replaces the quantum thermal average by an average over ring polymer
configurations on the extended phase space
$\R^{2dN} \times \{ 0, 1 \}^N$, which consists of not only the
position and momentum of the ring polymer, but also the the level
index of each bead. We will make precise the accuracy of the
approximation below in \S\ref{sec:derivation}.

Comparing \eqref{eq:reA} and \eqref{eq:reZ}, one observes that the
partition function \eqref{eq:reZ} under ring polymer representation
can be viewed as the thermal average with respect to the weight
function $W_N[I]$, where $I$ is the identity operator. Note that, the
kinks always show up in pairs, so a direct verification shows that
$W_N[I] =1$.

Next, we shall construct path integral molecular dynamics with surface
hopping (PIMD-SH) method to sample the thermal average based on the
above formulations in \S\ref{sec:pimdsh}. The derivation of the ring
polymer representation is given in \S\ref{sec:derivation}.

\subsection{PIMD-SH method}\label{sec:pimdsh}

We observe that \eqref{eq:reA} can be viewed as (up to a
normalization) an average with respect to the classical Gibbs
distribution for ring polymers on the extended phase space with
Hamiltonian $H_N$:
\begin{equation}\label{eq:ensembleavgA} \langle \wh{A} \rangle \approx
\frac{1}{(2\pi)^{dN}}\int_{\RR^{2dN}} \ud \bd q \ud \bd p
\sum_{\bd{\ell}\in\{0, 1\}^N} \pi (\widetilde {\bd{z}} )W_N[A]
(\widetilde {\bd{z}}),
\end{equation} with distribution
\begin{equation}\label{eq:pi} \pi (\widetilde {\bd{z}}) =
\frac{1}{\mathcal Z_N} \exp(-\beta_N H_N (\widetilde {\bd{z}})).
\end{equation} To simplify the notation, we have denoted by
$\widetilde {\bd z} = (\bd z, \bd\ell) \in \RR^{2dN} \times \{0,
1\}^N$ a state vector on the extended phase space, where $\bd z=(\bd
q, \bd p)$ are the position and momentum variables.  Notice that in
\eqref{eq:pi}, $\mc{Z}_N$ introduced in \eqref{eq:reZ} normalizes the
distribution in the sense that
\[ \frac{1}{(2\pi)^N}\int_{\RR^{2dN}} \ud \bd q \ud \bd p
\sum_{\bd{\ell}\in\{0, 1\}^N} \pi (\widetilde {\bd{z}} )=1.
\] As a result, if we can construct a trajectory $\widetilde
{\bd{z}}(t)$ that is ergodic with respect to the equilibrium
distribution $\pi$, we can sample the ensemble average on the right
hand side of \eqref{eq:ensembleavgA} by a time average to approximate
$\langle\wh{A}\rangle$:
\begin{equation} \langle \wh{A} \rangle \approx \lim_{T \rightarrow
\infty } \frac{1}{T} \int_0^T W_N [A] (\widetilde {\bd{z}} (t)) \ud t.
\end{equation} This is the basis of our path-integral molecular
dynamics with surface hopping (PIMD-SH) method.

The dynamics of $\wt{\bd{z}}(t)$ is constructed as follows. The
position and momentum part of the trajectory $\bd{z}(t)= (\bd q(t),
\bd p (t))$ evolves according to a Langevin dynamics with Hamiltonian
$H_N (\bd q (t), \bd p(t), \bd\ell (t))$ given the surface index
$\bd\ell (t)$, i.e., a Langevin thermostat is used.  
More specifically, we have
\[ \left\{
 \begin{array}{l l} \ud {\bd q} & = \nabla_{\bd p} H_N (\bd q (t), \bd
p(t), \bd\ell (t)) \ud t, \vspace{0.2cm} \\ \ud {\bd p} & =-
\nabla_{\bd q} H_N (\bd q (t), \bd p(t), \bd\ell (t)) \ud t
\vspace{0.1cm} \\ & \qquad\qquad - \gamma \bd p \ud t + \sqrt{2 \gamma
\beta^{-1}_N M} \ud \bd{B}.
 \end{array} \right.
\] 
Here $\bd{B}=\bd{B}(t)$ is a vector of $dN$ independent Brownian motion (thus
the derivative of each component is an independent white noise), and
$\gamma \in \mathbb R_+$ denotes the friction constant, as usual in
Langevin dynamics.  Notice that for
$\forall \, \bd\ell (t) \in \{ 0,1\}^N$,
\[ \nabla_{\bd p} H_N (\bd q (t), \bd p(t), \bd\ell (t))= \bd
p(t)/M.\] Thus the evolution of the position just follows as usual
\[ \dot{\bd{q}} = \frac{1}{M} \bd{p}.
\] The force term on the hand, given by $-\nabla_{\bd q} H_N (\bd q
(t), \bd p(t), \bd\ell (t))$, is in general $\bd\ell (t)$-dependent,
as the potential energy landscape depends on the level index $\bd\ell
(t)$. Hence, the evolution of the momentum $\bd{p}(t)$ depends on the
level index.

The evolution of $\bd{\ell}(t)$ follows a surface hopping type
dynamics in the spirit of the fewest switches surface hopping
\cite{Tully:90} (see also our recent works \cite{FGASH, FGASH2}). In
particular, we can take it to be a Markov jump process with
infinitesimal transition rate over the time period $(t, t + \delta t)$
for $\delta t \ll 1$ given by
\begin{multline}
  \mathbb{P} \bigl(\bd\ell (t+\delta t) = \bd\ell' \mid \bd\ell (t) = \bd\ell \, , \bd z(t)=\bd z\bigr)= \\
  = \delta_{\bd\ell' , \bd\ell} + \eta \lambda_{\bd\ell', \bd\ell}
  (\bd z) \delta t + o (\delta t).
\end{multline}
This means that if the current configuration of the ring polymer is
given by $(\bd{z}, \bd{\ell})$, during the time interval
$(t, t+\delta t)$, the level index might change to
$\bd{\ell}' \neq \bd{\ell}$ with probability
$\eta \lambda_{\bd{\ell}', \bd{\ell}}(\bd{z}) \delta t + o(\delta t)$. Here
$\eta>0$ is an overall scaling parameter for hopping intensity (the
larger $\eta$ is, the more frequent hopping occurs), the coefficients
$\lambda_{\bd\ell',\bd\ell}$ are specified as
\begin{equation} \label{eq:lambda}
  \lambda_{\bd\ell',\bd\ell}(\bd{z}) = 
  \begin{cases}
    \dps
    -  \sum_{\wt{\bd\ell} \in S_{\bd\ell}} p_{\wt{\bd\ell},\bd\ell}(\bd{z}), &  \bd\ell' =\bd\ell,  \\
    p_{\bd\ell',\bd\ell}(\bd{z}),   & \bd{\ell}' \in S_{\bd{\ell}}, \\
    0, & \text{otherwise},
\end{cases}
\end{equation}
where
$S_{\bd\ell}= \{ \bd{\ell}' \mid \norm{\bd\ell'- \bd\ell}_1=1 \;
\mbox{or} \; \bd\ell'= \bd 1 -\bd\ell \}$
denotes all allowed configuration $\bd{\ell}'$ after the hopping:
$\bd 1$ is the vector with all entries $1$, so
$\bd{\ell'} = \bd{1} - \bd{\ell}$ indicates that the surface index of
each bead is flipped; and
$\norm{\bd{\ell'} - \bd{\ell}}_1 = \sum_k \abs{\ell'_k - \ell_k} = 1$
indicates that one and only one bead jumps to the opposite energy
surface.  Here in the rate expression, $p_{\bd\ell',\bd\ell} (\bd z)$
is defined as
\[
p_{\bd\ell',\bd\ell} (\bd z) = \exp \left( \frac {\beta_N} 2 \bigl(
  H_N(\bd z, \bd\ell) - H_N(\bd z, \bd\ell') \bigr) \right), 
\] 
which is chosen so that the detailed balance relation is satisfied
\begin{multline}\label{eq:detailbalance}
  p_{\bd\ell',\bd\ell} (\bd z ) e^{- \beta_N H_N(\bd z, \bd\ell)}= 
  e^{- \frac {\beta_N} 2 \bigl(
    H_N(\bd z, \bd\ell) + H_N(\bd z, \bd\ell') \bigr)} \\
  = p_{\bd\ell,\bd\ell'} (\bd z ) e^{- \beta_N H_N(\bd z, \bd\ell')}.
\end{multline}
This guarantees that the distribution $\pi$ is preserved under the
dynamics of the jumping process (as will be further discussed below).

The above choice of $S_{\bd{\ell}}$ allows only two types of change of
level indices: either changing the surface index of one single bead
(single hop) or changing the surface index of all beads (total
flip). This is chosen for simplicity, as it ensures that any surface
index configuration can be reached and at the same time we do not need
to consider all possibilities at a single time step, which is
combinatorial and inefficient for practical implementation. 
 
To show that $\pi$ as in \eqref{eq:pi} is indeed the equilibrium
distribution corresponds to the dynamics of $\wt{\bd{z}}(t)$, it is
more convenient to write down the associated Fokker-Planck equation of
the dynamics.  Denote the probability distribution on the extended
phase space at time $t$ by $f(t, \widetilde {\bd{z}})$, $f$ satisfies
the following Fokker-Planck equation (\textit{i.e.}, forward Kolmogorov equation)
\begin{multline}\label{eq:FokkerPlanck}
  \frac{\partial}{\partial t} f (t, \bd z, \bd\ell) = \bigl\{H_N, f
  (t, \bd z,\bd\ell)\bigr\}_{\bd z} - \gamma \nabla_{\bd p} \cdot
  \Bigl(\bd p f +\frac{M}{\beta_N } \nabla_{\bd p} f \Bigr) \\+
  \sum_{\bd\ell'} \lambda_{\bd\ell,\bd\ell'} (\bd z) f(t, \bd
  z,\bd\ell'),
\end{multline}
where on the right hand side, the first term accounts for the
Hamiltonian part of the dynamics of $\bd{z}$, the second term comes
from the dissipation and fluctuation due to the Langevin thermostat,
and the last term is due to the jumping process of $\bd{\ell}$. In the
above equation, $\{ \cdot, \cdot \}_{\bd z} $ stands for the usual
Poisson bracket corresponding to the Hamiltonian dynamics
\[
\{h, f\}_{\bd z} = \nabla_{\bd q} h\cdot \nabla_{\bd p} f -
\nabla_{\bd p} h\cdot \nabla_{\bd q} f.
\]
Since the Boltzmann-Gibbs distribution is proportional to
$e^{-\beta_N H_N}$, which is a function of the Hamiltonian, it follows
that
\[
\bigl\{ H_N( \bd z, \bd\ell) , e^{- \beta_N H_N(\bd z, \bd\ell)} \bigr\}_{\bd{z}}=0.
\]
As usual for Langevin dynamics, the fluctuation-dissipation balance
ensures that
\[
\nabla_{\bd p} \cdot \Bigl(\bd p e^{- \beta_N H_N(\bd z, \bd\ell)} +
\frac{M}{\beta_N } \nabla_{\bd p} e^{- \beta_N H_N(\bd z, \bd\ell)}
\Bigr) =0.
\]
Moreover, using the detailed balance relation \eqref{eq:detailbalance}, we verify using the definition of  $\lambda$ as in \eqref{eq:lambda}  that
\begin{multline*}
\sum_{\bd\ell'} \lambda_{\bd\ell,\bd\ell'} (\bd z ) e^{- \beta_N H_N(\bd z, \bd\ell')} \\
= \sum_{\bd\ell' \in S_{\bd{\ell}}} p_{\bd{\ell}, \bd{\ell}'} e^{- \beta_N H_N(\bd z, \bd\ell')}  - \sum_{\bd\ell' \in S_{\bd{\ell}}} p_{\bd{\ell}', \bd{\ell}} e^{- \beta_N H_N(\bd z, \bd\ell)} 
=0.
\end{multline*}
Therefore, we conclude that the Boltzmann-Gibbs distribution, which is
a constant multiple of $e^{- \beta_N H_N(\bd z, \bd\ell)}$, is an
stationary solution to the Fokker-Planck equation. Note that this
remains the case regardless the choice of the hopping intensity
parameter $\eta$ and the friction parameter $\gamma$ (the latter is of
course a familiar fact for Langevin thermostat). We will study the
effect of tuning $\eta$ in our numerical examples in \S\ref{mutest}.

\subsection{Ring polymer representations for the two-level  Hamiltonians} \label{sec:derivation}

Let us now present the derivation of the ring polymer representation
for the two-level system, as presented in \S\ref{sec:ringpolymer}.  Recall the Hamiltonian in the diabatic
picture is given by
\[
\wh H = \frac{1}{2M}
\left[
\begin{array}{c c}
 \wh p^2  &  \\
  &   \wh p^2
\end{array}
\right] 
+
\left[
\begin{array}{c c}
 V_{00} (\wh q)  &    V_{01} (\wh q)  \\
 V_{10} (\wh q)  &    V_{11} (\wh q)
\end{array}
\right]
=: \wh T + \wh V .
\]
We recall here for convenience that we have assumed  $\wh V$ is real symmetric, and the off-diagonal function, which is denoted by $v$ for simplicity, does not change sign for $q \in \mathbb R^d$. We consider a large  fixed $N \in \mathbb N^+$ and introduce an
equispaced partition of $[0,\beta]$ (recall that
$\beta_N = \beta / N$),
\[
0< \beta_N < 2 \beta_N < \cdots < (N-1) \beta_N < N \beta_N = \beta. 
\]

By inserting resolution of identities with respect to position, we have for the
partition function 
\begin{equation}\label{eq:parmid}
\begin{aligned}
  \tr_{ne}[e^{-\beta \wh H}] & = \tr_{\C^2} \int_{\RR^d} \ud q_1 \bra{q_1} e^{-\beta \wh H} \ket{q_1}  \\
                             & = \tr_{\C^2} \int_{\RR^d} \ud q_1 \bra{q_1} e^{-\beta_N \wh H} \cdots e^{-\beta_N \wh H} \ket{q_1}  \\
                             & = \tr_{\C^2} \int_{\RR^{dN}} \ud \bd q  \prod_{k=1}^N \bra{q_k} e^{-\beta_N \wh H} \ket{q_{k+1}},
\end{aligned}
\end{equation}
where we have used the convention $q_{N+1}=q_1$ to simplify the
expression. So far, the reformulation is exact. Applying the Strang
splitting \cite{Strang:68} to the short imaginary time propagator
$e^{-\beta_N \wh H}$ and inserting a resolution of identity with
respect to momentum, we get
\begin{align*}
  & \bra{q_k} e^{-\beta_N \wh H} \ket{q_{k+1}} \\
  & = \bra{q_k} e^{-\beta_N \wh V / 2} e^ {- \beta_N \wh T } e^{-\beta_N
    \wh V / 2} \ket{q_{k+1}}  +\mathcal O (\beta_N^3) \\
  & = \int_{\RR^d}  \ud p_k \;  \bra{q_k} e^{-\beta_N \wh V / 2} \ket{p_k} \bra{p_k} e^ {- \beta_N \wh T } e^{-\beta_N
    \wh V / 2} \ket{q_{k+1}}\\ & \qquad  +\mathcal O (\beta_N^3). 
\end{align*}
Direct calculation of the right hand side leads to 
\begin{multline*}
  \bra{q_k} e^{-\beta_N \wh H} \ket{q_{k+1}} = \\
  = \frac{1}{(2 \pi)^d } \int_{\RR^d} \ud p_k \; e^{-\frac{\beta_N}{2}
    V(q_k)}e^{-\frac{\beta_N}{2} V(q_{k+1})}e^{-\beta_N S_{k}}
  + \Or(\beta_N^3),
\end{multline*}
where we suppress in the notation the dependence of $S_k$ on $p_k, q_k$ and $q_{k+1}$:
\begin{align*}
  S_{k}= \frac{p^2_k}{2M}+ \frac{M }{2 \beta_N^2} \left(q_k-q_{k+1} \right)^2 . 
\end{align*}
Note that, if the $V$ matrices are diagonal, we have recovered the
familiar terms in the usual ring polymer representation for single
energy surface. Let us emphasize though we consider the general case
that $V$ contains off-diagonal terms, and in particular,
$\exp(-\frac{\beta_N}{2} V)$ is a $2 \times 2$ matrix and hence
$\exp(-\frac{\beta_N}{2} V(q_k))$ in general does not commute with
$\exp(-\frac{\beta_N}{2} V(q_{k+1}))$.

Applying the above calculation to every term in the product on the
right hand side of \eqref{eq:parmid}, we obtain
\begin{multline}\label{eq:inter1}
\tr_{ne}[e^{-\beta \wh H}] =  \frac{1}{(2\pi)^{dN}}  \tr_{\C^{2}} \int_{\RR^{2dN}} \ud \bd q  \ud \bd p \;
e^{-\beta_N  V(q_1)  } \times  \\
\times e^{-\beta_N V(q_2)} \cdots e^{-\beta_N  V(q_N)  }  e^{-\beta_N \bigl(S_1+ \cdots S_N \bigr)}+\mathcal O(N \beta_N^3).
\end{multline}
We have an extra $N$ factor in the error $\Or(N \beta_N^3)$ since the
error from each operator splitting adds up. 

So far the basis for the discrete electronic states
  has not been fixed. In below, we will choose to use the diabatic
  picture, while the adiabatic picture can be also used, as the
  calculations shown in Appendix~\ref{App:C}. In the adiabatic
  picture, we recover the ring polymer representation of Schmidt and
  Tully in \cite{SchmidtTully2007}. The sampling strategy based on
  PIMD-SH can be applied to adiabatic picture as well, which we will
  leave for future studies.

Let us further simplify the integrand in the above equation in order
to arrive at the desired Boltzmann-Gibbs form as in \eqref{eq:reZ}. In
particular, to deal with the matrix exponential $e^{-\beta_N V}$, we
split the matrix potential $V$ into diagonal and off-diagonal parts
\[
 V=
 \begin{bmatrix}
   V_{00}   &    V_{01} \\
   V_{10} & V_{11}
 \end{bmatrix}
 = 
 \begin{bmatrix}
   V_{00}   &   0 \\
   0   &    V_{11} 
 \end{bmatrix}
 +
 \begin{bmatrix}
   0   &    V_{01} \\
   V_{10}   &    0 
 \end{bmatrix}
 =:  V_d +  V_o.
\]
Using another Strang splitting, we obtain 
\begin{equation}\label{eq:expV}
\begin{aligned}
  e^{-\beta_N V} & = e^{-\beta_N (V_d + V_o)} \\
                 & =  e^{-\beta_N V_d /2}e^{-\beta_N V_o }  e^{-\beta_N V_d /2}+ \mathcal O (\beta_N^3).   
\end{aligned}
\end{equation}
Explicit calculation gives the matrix exponential 
\begin{equation*}
  e^{-\beta_N V_o} = 
  \begin{bmatrix}
    \cosh(-\beta_N V_{01}) & \sinh(-\beta_N V_{01}) \\
    \sinh(-\beta_N V_{01}) & \cosh(-\beta_N V_{01}) 
  \end{bmatrix}. 
\end{equation*}
Substitute this into the right hand side of \eqref{eq:expV} and
rewrite the resulting matrix elements as exponentials, we arrive at 
\begin{multline}\label{eq:entryexpV}
  \bra{\ell} e^{-\beta_N V} \ket{\ell'}
  = \\ = \begin{cases}
    e^{-\beta_N V_{\ell \ell}+ \ln \bigl( \cosh(\beta_N \abs{V_{01}}) \bigr)} + \Or(\beta_N^3), 
    & \ell = \ell'; \\
    - \frac{V_{01}}{\abs{V_{01}}} e^{-\beta_N\frac{V_{00}+V_{11}}{2} + \ln\bigl( \sinh(\beta_N \abs{V_{01}}) \bigr) } + \Or(\beta_N^3), & \ell \neq \ell',
    \end{cases}
\end{multline}
where the prefactor $- V_{01} / \abs{V_{01}}$ when $\ell \neq \ell'$
is due to dependence of the sign of $\sinh(-\beta_N V_{01})$ on the
positivity of $V_{01}$. 

Applying the above expression for $e^{-\beta_N V}$ to 
\eqref{eq:inter1}, we arrive at the following approximation (recall
the definition of matrices $G_k$ in \eqref{eq:Gele})
\begin{multline*}
  \tr_{ne}[e^{-\beta \wh H} ] = \frac{1}{(2\pi)^{dN}} \int_{\RR^{2dN}}
  \ud \bd q \ud \bd p \, \sum_{\bd{\ell}\in \{ 0,1\}^N} \\ \times
  \exp\left( -\beta_N \sum_{k=1}^N \langle \ell_k | G_{k}| \ell_{k+1}
    \rangle \right) +\Or(N \beta_N^3), 
\end{multline*}
which justifies the ring polymer representation of the partition function \eqref{eq:reZ}. 
In order to get the above, we realize that the
$- V_{01} / \abs{V_{01}} $ factors for all kinks (for $k$ such that
$\ell_k \neq \ell_{k+1}$) in a ring polymer will cancel since there
are even number of kinks and $V_{01}$ does not change sign by
assumption (when $V_{01}$ is always negative, the factors are simply
always $1$; when $V_{01}$ is always positive, the factors are $-1$,
even number of them multiply to $1$). 

As a result, each term in the average \eqref{eq:reZ} is positive (as
it is an exponential), and thus we can view
$\exp(-\beta_N H_N(\bd{q}, \bd{p}, \bd{\ell}))$ as a probability
density for the ring polymer configuration
$(\bd{q}, \bd{p}, \bd{\ell})$. The PIMD-SH then samples this
distribution on the extended phase space.  This is no longer true
without the assumption that the off-diagonal entry of the matrix
potential $V_{01}$ does not change sign. While we can still take the
absolute value of the summand as the distribution, we would also need
to approximate the partition function by an average of terms which
change sign depending on the ring polymer configuration. The sign
change in general increases the difficulty of the sampling, this is a
manifestation of the familiar ``fermionic sign problem'' in quantum
Monte Carlo simulations, see e.g., \cite{Ceperley:10}.  Further
discussions on the formulation for a general two-level system can be found in the
Appendix~\ref{App:A} and will be explored in future works.

For an observable $\wh A$ that only depends on the position variable,
following a similar derivation leads to \eqref{eq:inter1}, we have
\begin{multline}
  \tr_{ne} [e^{-\beta \wh H} \wh A] = \frac{1}{(2\pi)^{dN}}
  \tr_{\C^{2}} \int_{\RR^{2dN}} \ud \bd q \ud \bd p \;
  A(q_1) \times    \\
  \times e^{-\beta_N V(q_1) } e^{-\beta_N V(q_2)} \cdots e^{-\beta_N
    V(q_N) } \\
  \times e^{-\beta_N (S_1+ \cdots S_N )} + \mathcal O(N \beta_N^3).
\end{multline}
By symmetry, we can also move the $A$ matrix to before
$e^{-\beta_N V(q_k)}$ and evaluate $A$ at $q_k$. Taking an average
over all the possibilities, we get
\begin{multline}
  \tr_{ne} [e^{-\beta \wh H} \wh A] = \frac{1}{N} \sum_{k=1}^N
  \frac{1}{(2\pi)^{dN}} \tr_{\C^{2}} \int_{\RR^{2dN}} \ud \bd q \ud
  \bd p \;
  \\
  \times e^{-\beta_N V(q_1) } \cdots e^{-\beta_N V(q_{k-1})} A(q_k) \\
  \times
  e^{-\beta_N V(q_k)} \cdots e^{-\beta_N V(q_N)} \\
  \times e^{-\beta_N (S_1+ \cdots S_N )} +\mathcal O(N
  \beta_N^3).
\end{multline}
Again using the expansion \eqref{eq:expV} of $e^{-\beta_N V}$, we
arrive at (recall that $\bar{\ell}_k = 1 - \bar{\ell}_k$)
\begin{widetext}
\begin{equation*}
  \begin{aligned}
    \tr_{ne}[e^{-\beta \wh H} \wh A] & = \frac{1}{N} \sum_{k=1}^N
    \frac{1}{(2\pi)^{dN}} \int_{\RR^{2dN}} \ud \bd q \ud \bd p
    \sum_{\bd{\ell}\in\{0, 1\}^N} \exp\Bigl( -\beta_N \sum_{k'=1}^N
    \bra{\ell_{k'}} G_{k'} \ket{\ell_{k'+1}} \Bigr) \times \\
    & \quad \times \sum_{\ell'=0,1} e^{\beta_N \langle \ell_k| G_k|
      \ell_{k+1} \rangle - \beta_N \bra{\ell'} G_{k} \ket{ \ell_{k+1}}}
 \bra{\ell_{k}} {A( q_{k})} \ket{\ell'} \; 
\bra{\ell'} J_{k}
    \ket{\ell_{k+1}} \prod_{\tilde k\ne k} \bra{\ell_{\tilde k}}
    J_{\tilde k} \ket{\ell_{\tilde k+1}} +\Or (N \beta_N^3) \\
    & = \frac{1}{N} \sum_{k=1}^N \frac{1}{(2\pi)^{dN}}
    \int_{\RR^{2dN}} \ud \bd q \ud \bd p \sum_{\bd{\ell}\in\{0, 1\}^N}
    \exp\Bigl( -\beta_N \sum_{k'=1}^N
    \bra{\ell_{k'}} G_{k'} \ket{\ell_{k'+1}} \Bigr) \times \\
    & \quad\times \biggl( \bra{\ell_k} A(q_k) \ket{\ell_k} -
    e^{\beta_N \langle \ell_k| G_k| \ell_{k+1} \rangle - \beta_N
      \bra{\bar{\ell}_k} G_{k} \ket{ \ell_{k+1}}}  \bra{\ell_{k}}
  A(q_{k}) \ket{\bar{\ell}_k} \frac{V_{\ell_k\bar{\ell}_k}}{\abs{V_{\ell_k\bar{\ell}_k}}}\biggr) + \Or (N \beta_N^3)
\end{aligned}
\end{equation*}
\end{widetext}
where in the first equality, we have used the short hand notations $J_k$:
\begin{equation}\label{eq:Jele}
\bra{\ell} J_{k} \ket{\ell'} =
\begin{cases}
   1,  &  \ell=\ell', \\
   - \frac{V_{\ell\ell'}(q_k)}{\abs{V_{\ell\ell'}(q_k)}}, & \ell\neq
   \ell'.
 \end{cases}
\end{equation}
The above approximation of $\tr_{ne}[e^{-\beta \wh H} \wh A]$ is exactly \eqref{eq:reA} by checking the definition of the
weight function in \eqref{eq:WNA}.

\section{Time integrator for PIMD-SH} \label{sec:algorithm}

Recall that given the two level quantum Hamiltonian $\wh H$ and
$\beta$, we choose a sufficiently large $N\in \mathbb N$, and the
effective Hamiltonian for the ring polymer representation $H_N$ is
given in \eqref{eq:Ham}. The PIMD-SH dynamics is ergodic with respect
to the corresponding Gibbs distribution of the Hamiltonian. In
practical implementations, the PIMD-SH dynamics is discretized in time
to sample the equilibrium distribution.

To start with, we specify initial conditions to the sampling
trajectory
$\widetilde {\bd{z}} (0)=(\bd q (0), \bd p(0), \bd\ell(0))$. Due to
the ergodicity of the dynamics, any initial conditions can be used,
while a better initial sampling will accelerate the convergence of the
sampling. In our current implementation, for simplicity, we initialize
all the beads in the same position, sample their momentum by the
Gaussian distribution $\mathcal N (0, M \beta_N^{-1})$, and take
$\bd\ell(0) = \bd{0}$, where $\bd{0}$ is a vector of zeros, meaning
that initially all beads of the ring polymer stay on the lower energy
surface. Possible better initial sampling strategies can also be used.

The overall strategy we take for the time integration is time
splitting schemes, by carrying out the jumping step, denoted by J, and
the Langevin step denoted by L, in an alternating way.  In this work, 
we apply the Strang splitting, such that the resulting splitting
scheme is represented by JLJ. This means that, within the time interval
$[ t^n, t^n +\Delta t]$ ($\Delta t$ being the time step size), we
carry out the following steps in order:
\begin{enumerate}
\item We numerically simulate the jumping process for $\bd{\ell}$ for
  $\Delta t /2$ time with fixed position and momentum of the ring polymer;

\item We propagate numerically the position and momentum of the ring
  polymer using a discretization of the Langevin dynamics for
  $\Delta t$ time while fixing the surface index $\bd\ell$ (from the previous sub-step);

\item The jumping process for $\bd{\ell}$ is simulated for another $\Delta t/2$ time
  with fixed position and momentum of the ring polymer;

\item The weight function $W_N[A](\widetilde {\bd{z}} (t^{n+1}))$ of
  the observable $\wh{A}$ is calculated (and stored, if needed) to update the
  running average of the observable.
\end{enumerate}
The above procedure is repeated for each time step until we reach a
prescribed total sampling time $T$ or when the convergence of the
sampling is achieved under certain stopping criteria.  In this work,
we use a standard Monte Carlo scheme for the jumping process and the
BAOAB integrator for the Langevin dynamics, the details of both will
be further elaborated in subsections.

We remark that from a numerical analysis point of view, the above
splitting scheme corresponds to a splitting of the Fokker-Planck
equation introduced in \eqref{eq:FokkerPlanck}: The jump process step corresponds to 
\[
\frac{\partial}{\partial t} f (t, \bd z, \bd\ell)=\sum_{\bd\ell'} \lambda_{\bd\ell,\bd\ell'} (\bd z) f(t, \bd z,\bd\ell'),
\] 
while the Langevin step corresponds to 
\[
\frac{\partial}{\partial t} f (t, \bd z, \bd\ell) = \{ H_N, f (t, \bd  z,\bd\ell)\}_{\bd z} - \gamma \nabla_{\bd p} \cdot \left(\bd p f + \frac{M}{\beta_N } \nabla_{\bd p} f \right).
\] 
In particular, this leads to error analysis of the weak order of the
proposed scheme. We shall not go into the details of numerical
analysis in this work.

In what follows, we present the details of each steps in the splitting scheme.

\subsection{Simulation of the jumping process} \label{jump} 

Within a short time interval $\Delta t$, we consider the possible
jumps of the surface index $\bd{\ell}$, while fixing the position and
momentum of the ring polymer. For simplicity, we suppress the
appearance of $\bd q$ and $\bd p$ in the notation in this subsection,
as they do not change.

We assume that the time interval is so short that we may only consider
one or no jump during the interval, \textit{i.e.,}, two jumps
happening at probability $\Or(\Delta t)^2$ can be neglected.  Recall
the hopping intensity \eqref{eq:lambda} for the jumping process, in
particular, by our choice of $S_{\bd{\ell}}$, only jumps with
$\norm{\bd\ell-\bd\ell'}_1 = 1$ or when $\bd\ell = \bd 1 -\bd\ell'$
are allowed. Thus, the probability of a jump occurs from the current
level index $\bd{\ell}$ to $\bd{\ell'} \neq \bd{\ell}$ during the
time interval $\Delta t$ is given by
\begin{equation}
  h_{\bd\ell', \bd\ell} = \begin{cases}
    \Delta t \lambda_{\bd{\ell}', \bd{\ell}} = \eta \Delta t  p_{\bd\ell', \bd\ell},  &  \bd{\ell}' \in S_{\bd{\ell}};\\
    0, & \text{otherwise},
\end{cases}
\end{equation}
where recall that
$S_{\bd\ell}= \{ \bd{\ell}' \mid \norm{\bd\ell'- \bd\ell}_1=1 \;
\mbox{or} \; \bd\ell'= \bd 1 -\bd\ell \}$.
We assume that $\Delta t$ is chosen sufficiently small that
\begin{equation}\label{eq:conddeltat}
  \sum_{\bd\ell' \neq \bd\ell} h_{\bd\ell', \bd\ell} \leq 1,
\end{equation} 
and the probability that the level index is unchanged is then
$1 - \sum_{\bd\ell' \neq \bd\ell} h_{\bd\ell', \bd\ell}$. Thus a
uniform random number on $[0, 1]$ can be drawn to decide which event
happens (as in standard kinetic Monte Carlo simulations).

Recall that the choice of the hopping intensity parameter $\eta$ does
not change the equilibrium distribution of the PIMD-SH. In practical
simulations, one can adjust the parameter $\eta$ to make surface
hoppings more often, which in certain cases help accelerate the
sampling. This is particularly useful when the observable has
off-diagonal elements. We shall numerically verify the effect the this
parameter in \S\ref{mutest}. Of course, a larger $\eta$ requires
smaller time step size to ensure the condition \eqref{eq:conddeltat}
and also for accuracy. Therefore, direct simulation using a very large
$\eta$ might be difficult. On the other hand, we can explore similar
strategies as in the infinite swapping replica exchange molecular
dynamics \cite{LuVandenEijnden:13, YuLuAbramsVE:16} to simulate the
limiting system directly. This will be left for future works.


\subsection{The BAOAB scheme for Langevin dynamics} \label{sec:BAOAB}

With a fixed level index $\bd{\ell}$, the Langevin step then
corresponds to the propagation of the position and momentum
$(\bd q, \bd p)$ according to
\[
\left\{
 \begin{array}{l l}
\ud {\bd q} &  = {M}^{-1} {\bd p} \ud t, \\
\ud {\bd p} &   =-  \nabla_{\bd q} H_N \ud t - \gamma \bd p \ud t + \sqrt{2 \gamma \beta^{-1}_N}M^{1/2} \ud W .
 \end{array}
 \right.
\]
To numerically integrate the Langevin dynamics, we apply the BAOAB
splitting scheme \cite{LeimkuhlerMatthews} developed in the context of
Langevin thermostat for classical molecular
dynamics. The BAOAB scheme has been applied to and
  analyzed for conventional PIMD simulations (for single energy
  surface) recently and exhibit advantageous numerical performance
  over other numerical integrators \cite{LiuLiLiu:16} \footnote{The
    work \cite{LiuLiLiu:16} actually considered two slightly different
    variants of BAOAB style splitting, the BAOAB splitting we use
    here, which is consistent with \cite{LeimkuhlerMatthews},
    corresponds to what \cite{LiuLiLiu:16} referred as
    BAOAB-num.}. In the BAOAB scheme, the Langevin dynamics is
divided into three parts, the kinetic part (denoted by ``A''),
\[
\left\{
 \begin{array}{l l}
\ud {\bd q} &  = {M}^{-1} {\bd p} \ud t, \\
\ud {\bd p} &   =0 ,
 \end{array}
 \right.
\]
the potential part (denoted by ``B'')
\[
\left\{
 \begin{array}{l l}
\ud {\bd q} &  = 0, \\
\ud {\bd p} &   =-  \nabla_{\bd q} H_N \ud t ,
 \end{array}
 \right.
\]
and the Langevin thermostat part (denoted by ``O'')
\[
\left\{
 \begin{array}{l l}
\ud {\bd q} &  =0, \\
\ud {\bd p} &   =- \gamma \bd p \ud t + \sqrt{2 \gamma \beta^{-1}_N}M^{1/2} \ud W .
 \end{array}
 \right.
\]
The nice feature of such splitting is that each of these substeps can
be integrated explicitly. For example, the O part has the
following exact solution in the sense of the generated probability
distributions
\[
\left\{
 \begin{array}{l l}
 {\bd q}(t) &  = {\bd q}(0) , \\
 {\bd p}(t) &   =e^{- \gamma t}\bd p (0) + \sqrt{(1- e^{-2 \gamma t})(\beta_N^{-1}M)} \bd{n},
 \end{array}
 \right.
\]
where each component of $\bd{n}$ is an independent standard Gaussian random variable. 

The BAOAB scheme stands for a splitting scheme solving B part and A
part for a half time step $\Delta t/2$, followed by solving O part for
a full time step $\Delta t$, and followed by solving A part and B part
for another half time step $\Delta t /2$. Each substep uses the exact
time propagations.  Compared with other prevailing splitting methods
for the Lagevin dynamics, the BAOAB scheme has demonstrated higher
accuracy without sacrificing computational efficiency. Also, it is
also demonstrated \cite{LeimkuhlerMatthews} the BAOAB scheme enables
using larger time steps in the simulation while keeping the stability
of the integrator. 

\smallskip 

Combining the BAOAB splitting scheme with the step of jumping process,
our overall scheme can be represented as JBAOABJ. While other
numerical integrators are possible, in our numerical experiments shown
in Section~\ref{sec:test}, the JBAOABJ scheme seems to perform quite
well for the PIMD-SH sampling.

\section{Numerical results} \label{sec:test}

\subsection{Test examples} \label{sec:testex} 
\begin{figure}[ht]
\includegraphics[scale=0.55]{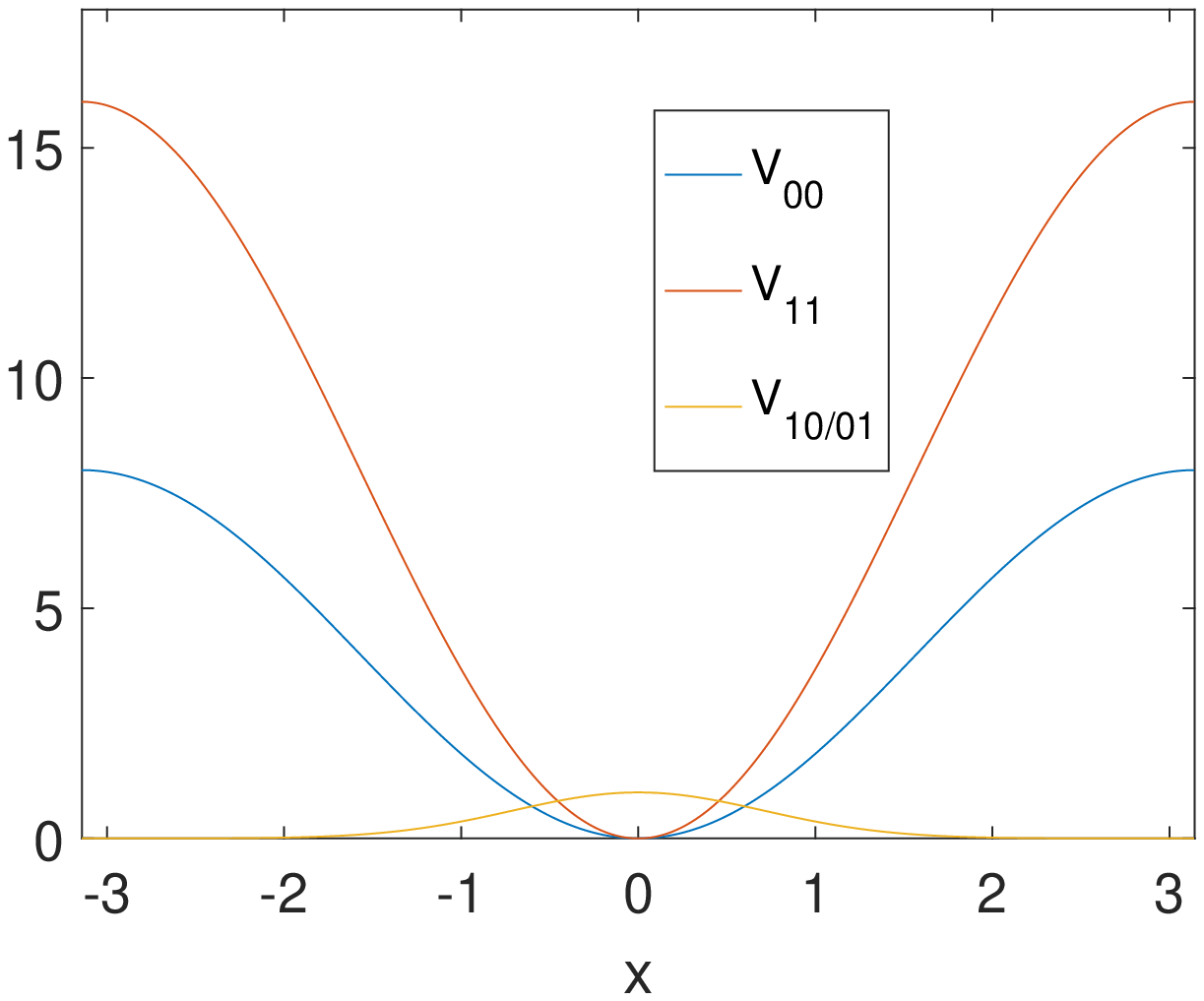} \\
\includegraphics[scale=0.55]{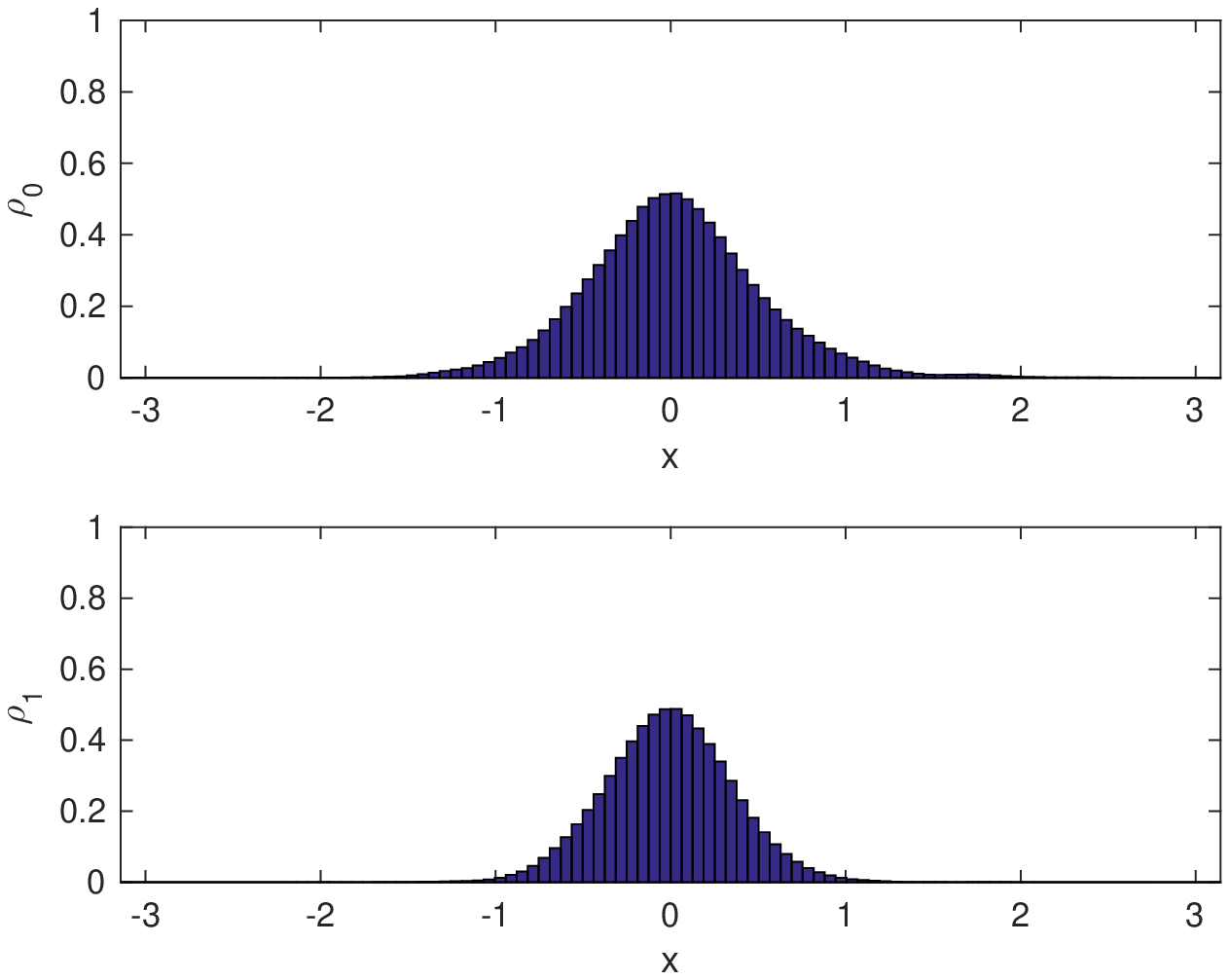}\\
\caption{Top: diabatic potential surfaces for the test example
  \eqref{ex:pot1}. Bottom: equilibrium distribution on both
  surfaces. }
\label{fig:Eplot1}
\end{figure}

To validate the PIMD-SH
method and to understand the choice of parameters in the method, we
test with following two potentials. Both potentials are chosen to be
one-dimensional and periodic over $[-\pi,\pi]$, so that the reference
solutions can be obtained accurately with pseudo-spectral
approximations and compared to the PIMD-SH results. The first test
potential is given
by 
\begin{equation}  \label{ex:pot1}
\left \{
\begin{split}
V_{00} &=4 \bigl(1-\cos(x) \bigr); \\
V_{11} & = 8\bigl(1-\cos(x) \bigr);\\
V_{01} & = V_{10} =  e^{-x^2}.
\end{split}
\right.
\end{equation}
Clearly, $V_{11} \ge V_{00}$ and the two energy surfaces only
intersect at $x=0$, where the off-diagonal term takes its largest
value.  
The energy surfaces are symmetric with respect to $x = 0$. At thermal
equilibrium, the density are expected to concentrate around $x=0$,
where transition between the two bands is the most noticeable due to
the larger off-diagonal coupling terms.  The diabatic
energy surfaces are plotted in Figure \ref{fig:Eplot1}(top).
Moreover, for $\beta=1$, $M=10$ $N=64$, $\Delta t = 1\mathrm e-3$ and
$T=1\mathrm e4$, we plot the (numerically obtained) marginal
equilibrium distribution of position variable on each energy surface
in Figure \ref{fig:Eplot1}(bottom).

\begin{figure}[ht]
\includegraphics[scale=0.55]{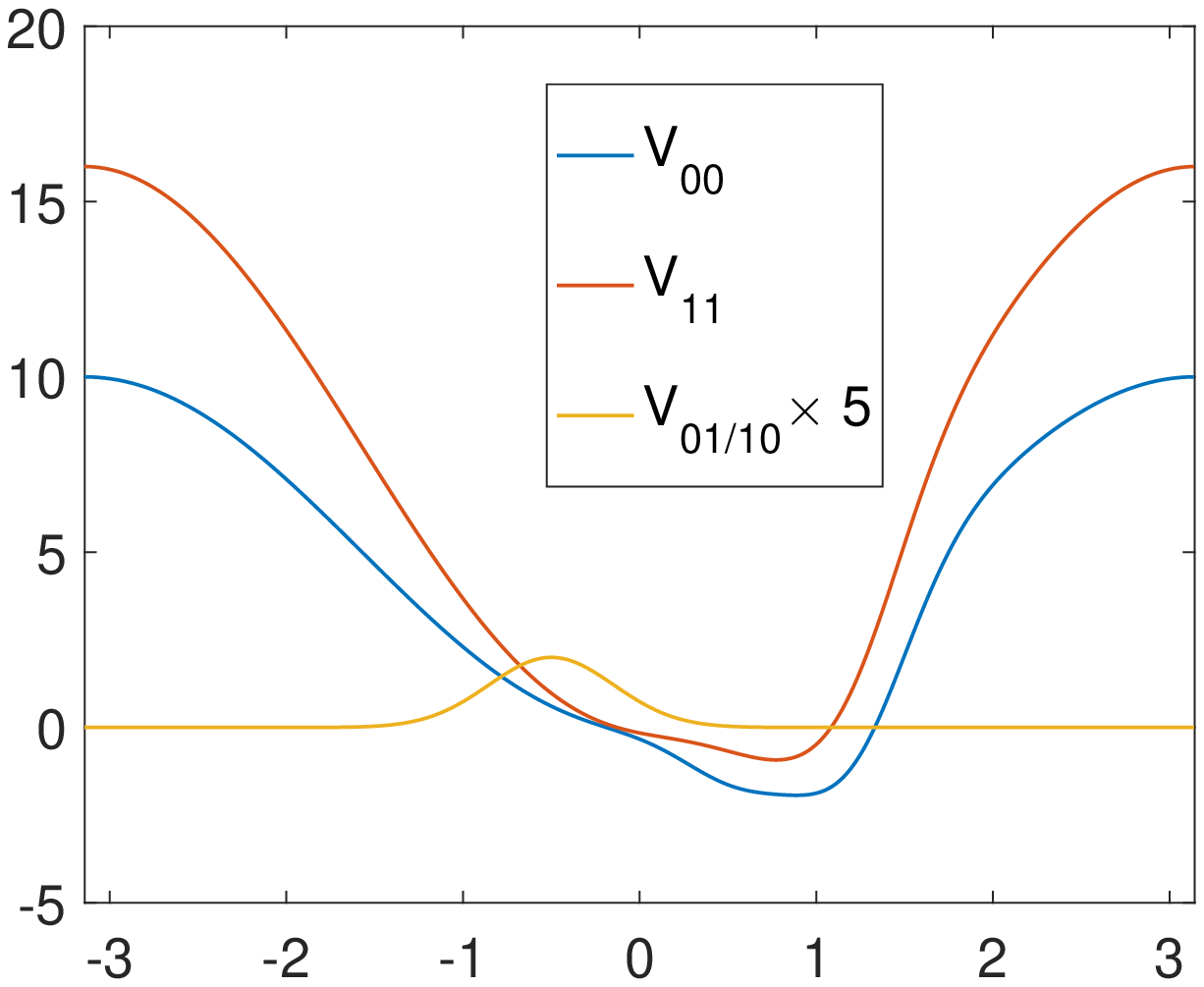} \\
\includegraphics[scale=0.55]{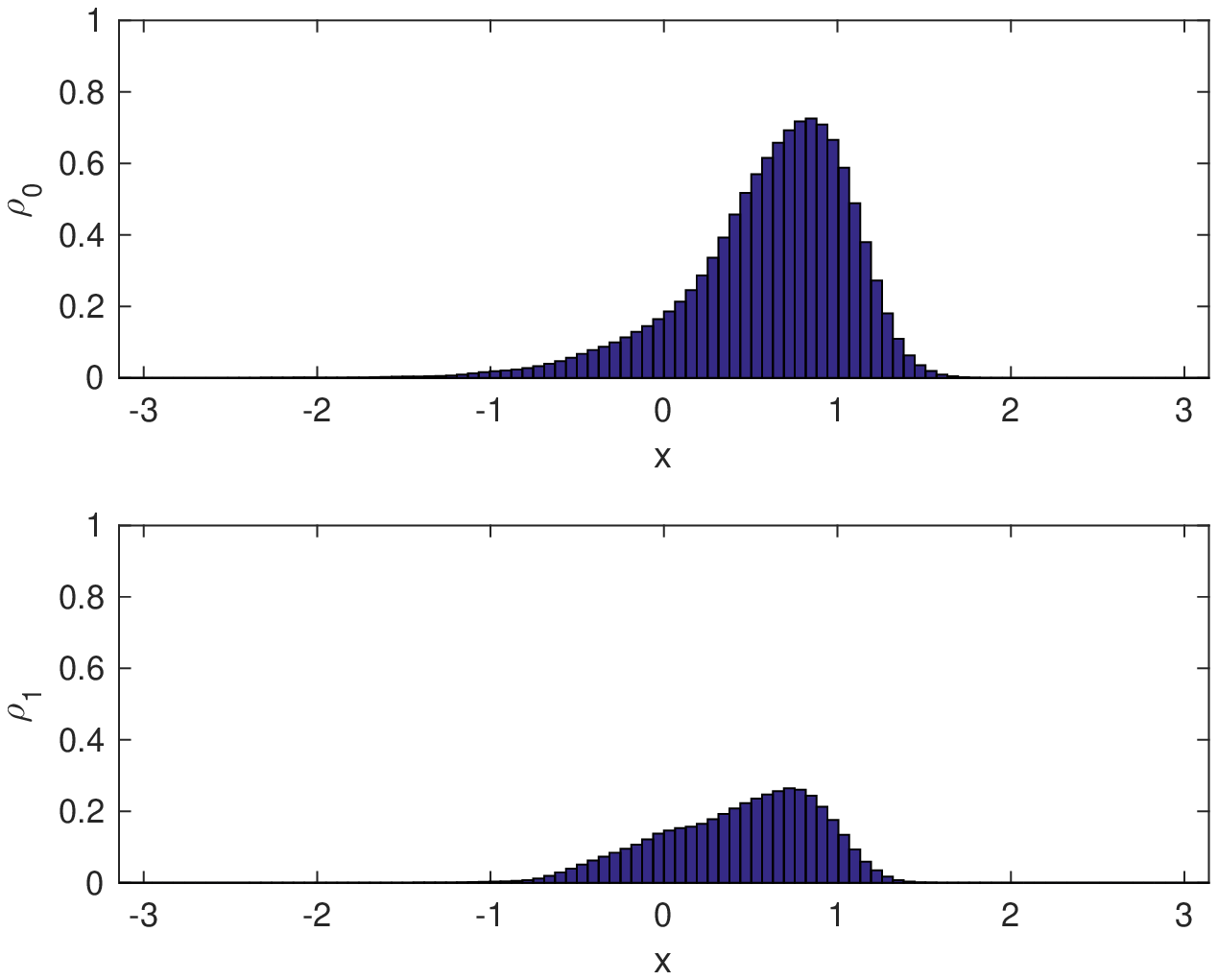}\\
\caption{Top: diabatic potential surfaces for the test example
  \eqref{ex:pot2}. Bottom: equilibrium distribution on both surfaces.}
\label{fig:Eplot2}
\end{figure}

The other test potential we take is given by 
\begin{equation}  \label{ex:pot2}
\left \{
\begin{split}
V_{00} &= 5-5 \cos(x)-4 e^{-5(x-1.2)^2}-2 e^{-5(x-0.6)^2}; \\
V_{11} & =8-8\cos(x)-3e^{-5(x-1.2)^2}-2e^{-4(x-0.8)^2};\\
V_{01} & = V_{10} = 0.4 e^{-4(x+0.5)^2}.
\end{split}
\right.
\end{equation}
The potential is designed so that $V_{11} \ge V_{00}$, and the two
energy surfaces achieve their minima around $x=1$.  Moreover, the
energy surfaces almost intersect when $x$ is slightly less than $0$,
and the off-diagonal potential is most noticeable around
$x=-0.5$. Thus, in this model, the potential is asymmetric, and the
location where the equilibrium distribution is concentrated deviates
from the most active hopping area. These make this test model more
challenging than the previous one. We plot the diabatic energy
surfaces in Figure \ref{fig:Eplot2}(top); and for $\beta=1$, $M=10$,
$N=64$, $\Delta t = 0.001$ and $T=10000$, we plot the (numerically obtained) marginal equilibrium distribution on each energy surface in Figure \ref{fig:Eplot2}(bottom).

Note that in both test cases, the off-diagonal potential $V_{01}$ has the same sign for all $q$, as we have assumed for the proposed method in this work. 
We will study  the cases when the observables are diagonal, or only its off-diagonal elements are nonzero.  When, the observables are diagonal, i.e., $\langle \ell | A | \ell' \rangle =0$, when $\ell \neq \ell'$, the weight function of the observable simplifies to 
\begin{equation*}
  W_N[A]  (\bd q, \bd p, \bd \ell) = 
  \frac{1}{N} \sum_{k=1}^N   
  \langle \ell_{k}|{A( q_{k})}| \ell_{k} \rangle.
\end{equation*}
From the simplified representation of the weight function, we learn that each beads gives similar contribution to the observations. 

When the diagonal elements of the observable are zero, i.e.,
$\langle \ell | A | \ell \rangle =0$ for $\ell = 0, 1$, the weight
function of the observable simplifies to
\begin{multline*}
W_N[A]  (\bd q, \bd p, \bd \ell) 
=  -\frac{1}{N} \sum_{k=1}^N  e^{\beta_N \langle \ell_{k}| G_k| \ell_{k+1} \rangle-\beta_N \langle \bar{\ell}_{k}| G_{k} | \ell_{k+1} \rangle} \\
\times 
\langle \ell_{k}|{A( q_{k})}|  \bar{\ell}_{k} \rangle \frac{V_{\ell_k\bar{\ell}_k}}{\abs{V_{\ell_k\bar{\ell}_k}}}. 
\end{multline*}
From which we see that, due to the exponential factor in the weight
function, the kinks of a ring polymer might give larger contributions
to the observables compared to other beads. The uneven contribution to
the observable from beads brings in numerical challenges since the
sample variance may be large. Numerical study of related issues is
presented in \S\ref{mutest}.

\subsection{Convergence with number of beads}

\begin{figure}[ht]
\centering
\includegraphics[scale=0.6]{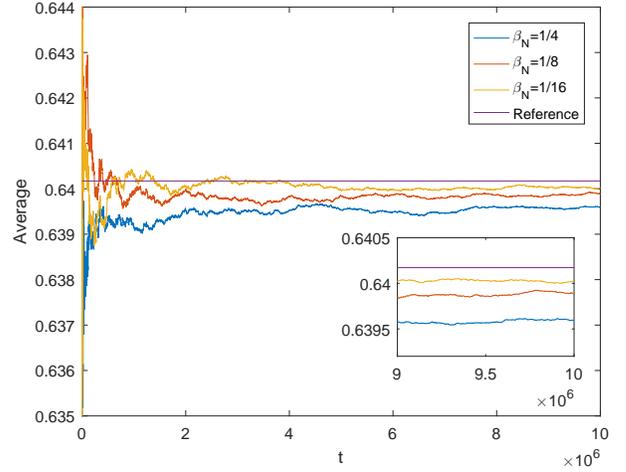}
\caption{Test potential \eqref{ex:pot2}, running average of the diagonal observable with different $\beta_N$.}
\label{fig:betaN1}
\end{figure}

\begin{figure}[ht]
\centering 
\includegraphics[scale=0.6]{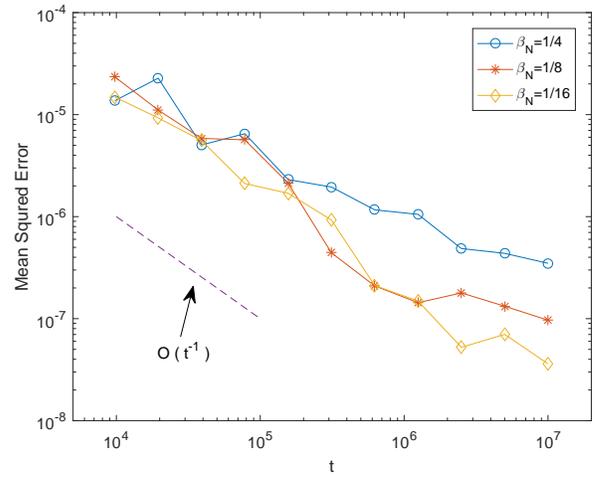}
\caption{Test potential \eqref{ex:pot2}, snapshots of mean squared
  error with $\beta_N=\frac{1}{4}$, $\frac 1 8$ and $\frac{1}{16}$ as a function of simulation time.}
\label{fig:time}
\end{figure}

We first test the PIMD-SH method for the Hamiltonian with the test
potential \eqref{ex:pot2} with $\beta=1$, $M=10$ and $\eta=1$. We
carry out the simulations with the following diagonal observable
\begin{equation} \label{eq:ob1}
A=\left[
\begin{array}{c c}
  e^{- {\wh q}^2}   &  0\\
 0 &   e^{- {\wh q}^2}  
\end{array}
\right]
\end{equation}
for $\beta_N=\frac{1}{4}$, $\frac{1}{8}$ and $\frac{1}{16}$ (and
correspondingly different number of beads in the ring polymer),
$\Delta t = 1\rm e-3$ and $T=1\rm e7$. The time step size is chosen
sufficiently small to ensure the accuracy of numerical integration. We
plot the running average of each simulation in
Figure~\ref{fig:betaN1}, from which we observe that the running
average for each $\beta_N$ approaches a steady value and is close to
the reference solution. We further compute the mean squared errors for
different simulation time for $\beta_N=\frac{1}{4}$, $\frac 1 8$ and
$\frac{1}{16}$, and plot them in Figure~\ref{fig:time}, from which we
observe that all three tests behave similarly, and the mean squared
errors decay roughly in proportion to $\mathcal O(t^{-1})$ until
around $t=1\rm e5$, consistent with the convergence rate of the Monte
Carlo sampling error. The asymptotic errors become noticeable when
$t = 1\rm e5 \sim 1\rm e7$, which tell apart the different $\beta_N$
with also reduced rates for decrease of the error.


The errors in empirical averages together with the $95 \%$ confidence
intervals and the mean squared errors are shown in Table
\ref{table:test1_2}, from which we see that the result approximates
the correct expectation of the observable while both the sampling
error and the asymptotic error make contributions to the total
numerical error. The mean squared errors of the empirical averages are
defined as $\text{M.S.E.}=\text{Bias}^2+ \text{Var}$, where
$\text{Bias}$ is calculated using the reference value and $\text{Var}$
is estimated using the observed data and the effective sampling
size. The result further confirms that increasing number of beads help
reduce the error in the PIMD-SH method.

The test for different $\beta_N$ is also carried out for the first potential, though the conclusion is pretty similar (and the second potential is more challenging), and hence we omit these results here.

\begin{table}
  \centering

  \begin{tabular}{ c| c| c|c} \hline
    & $\beta_N=\frac{1}{4}$ & $\beta_N=\frac{1}{8}$ & $\beta_N=\frac{1}{16}$    \\ \hline
Error &    5.80e-04 & 2.89e-04 & 1.52e-04
    \\ \hline
   95\% C.I. &2.23e-04 & 2.23e-04 & 2.24e-04
    \\ \hline
M.S.E. & 3.49e-07 &    9.67e-08 &  3.61e-08  \\ \hline
\end{tabular}
\caption{Errors in numerical empirical averages with 95 \% confidence intervals and mean squared errors. The reference value is $0.640172$. }
  \label{table:test1_2}

\end{table}

\subsection{Convergence with different time step sizes}
\begin{figure}[ht]
\centering
\includegraphics[scale=0.6]{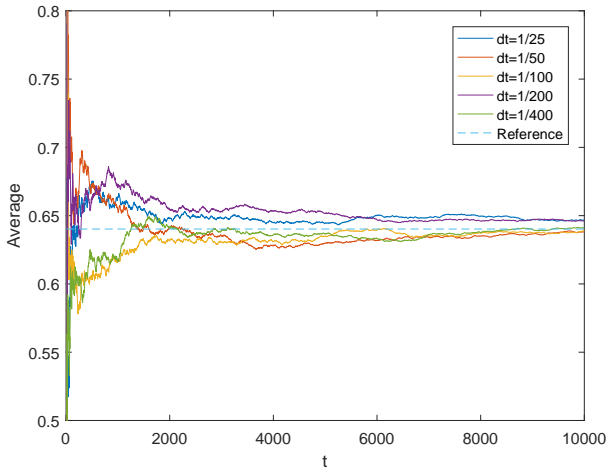}
\caption{Running average of a diagonal observable with test
  potential~\ref{ex:pot2} for different time step size $\Delta t$.}
\label{fig:diffdt}
\end{figure}

\begin{figure}[ht]
\begin{center}
\includegraphics[scale=0.6]{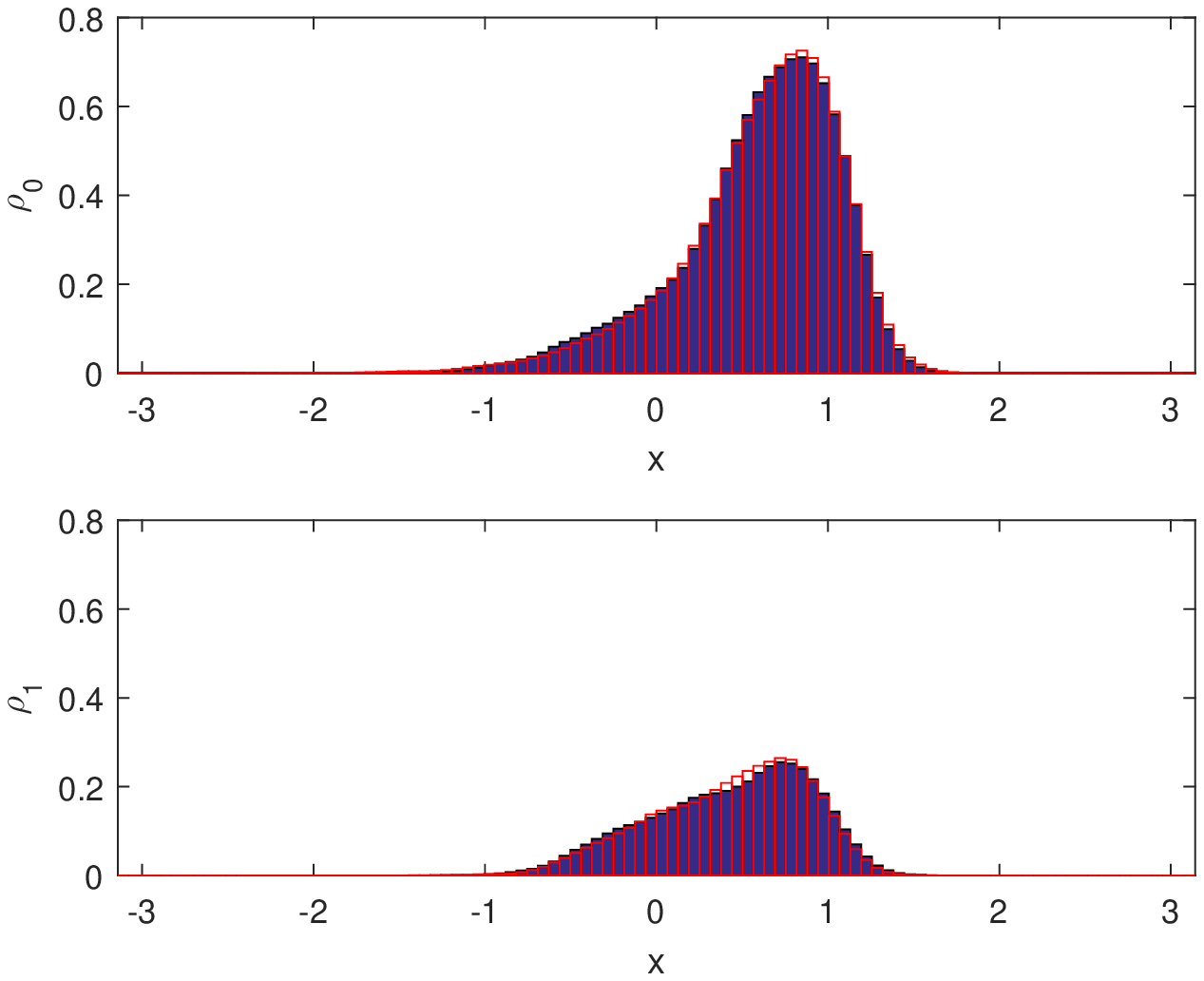} \\
\includegraphics[scale=0.6]{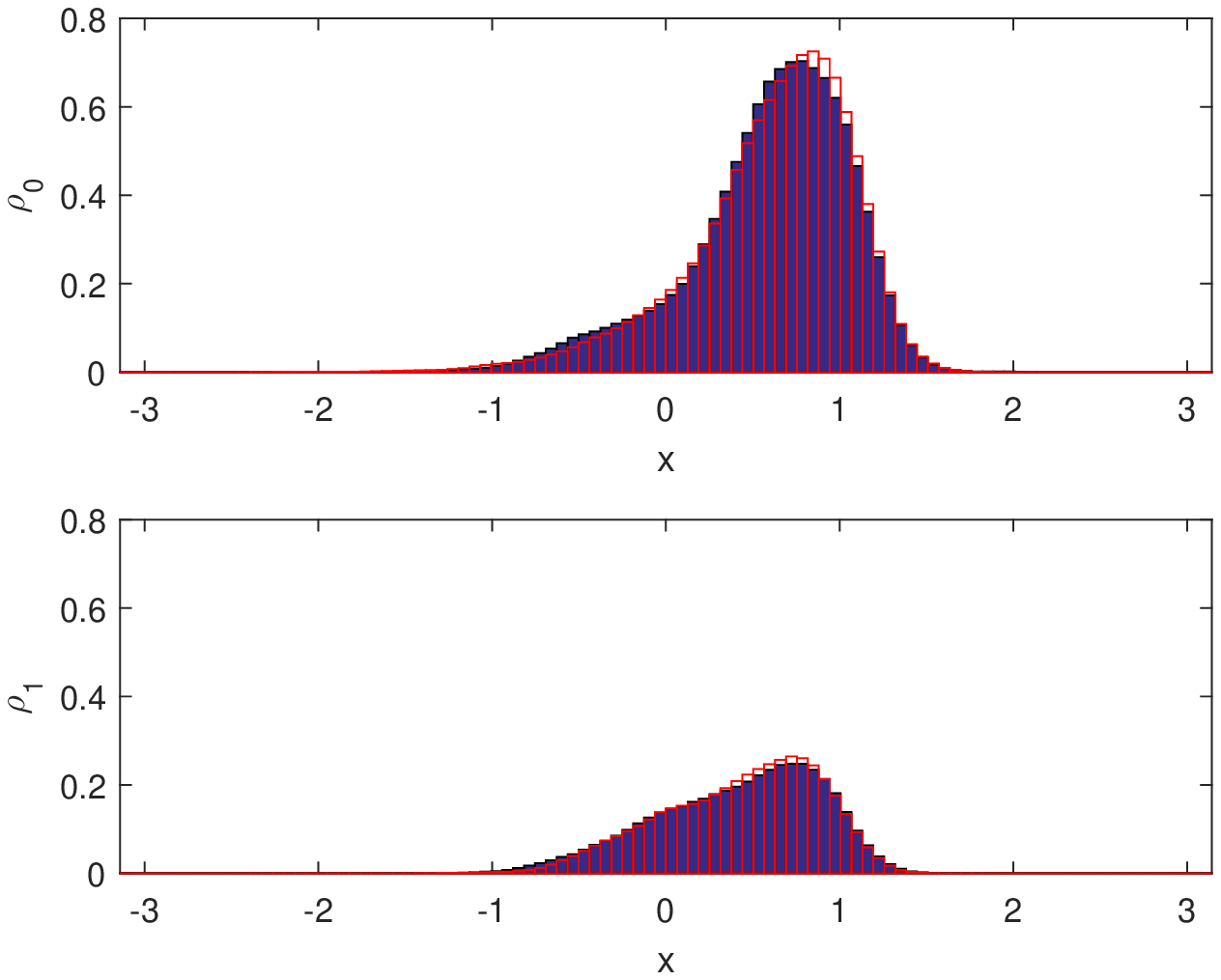}
\end{center}
\caption{Empirical histogram (blue) for the spatial distribution on the two
  energy surfaces with: $\Delta t= \frac{1}{25}$ (top) and
  $\Delta t = \frac{1}{400}$ (bottom), the reference solution computed
  in Section \ref{sec:testex} is plotted in
  red. 
}
\label{fig:diffdtconf}
\end{figure}

In this test, we use the test potential \eqref{ex:pot2}, with
$\beta=1$, $\beta_N=\frac{1}{16}$, $M=10$ and $\eta=1$, and vary the
time step sizes for the numerical integration. We choose the same
diagonal observable as before in \eqref{eq:ob1}.  The PIMD-SH method
is tested with time step sizes $\Delta t = \frac{1}{25}$,
$\frac{1}{50}$, $\frac{1}{100}$, $\frac{1}{200}$ and $\frac{1}{400}$
with total simulation time $T=10000$. The running averages of the
sampling for various parameters are plotted in
Figure~\ref{fig:diffdt}. The errors in the empirical averages together
with their $95\%$ confidence intervals and mean squared errors are
shown in Table~\ref{table:test2_1}. We observe that the PIMD-SH
captures the correct thermal average of the observable even for a
relatively large $\Delta t$. While the MSE decays for smaller time
step, the decrease is not too significant, as the sampling error is
probably dominant.  We also plot the empirical histogram for the
spatial distribution on each energy level in Figure
\ref{fig:diffdtconf} for two choices of the time step sizes.  

\begin{table}
  \centering
  \begin{tabular}{ c| c| c|c|c|c} \hline
    & $\Delta t=\frac{1}{25}$ & $\Delta t=\frac{1}{50}$ & $\Delta t=\frac{1}{100}$ & $\Delta t=\frac{1}{200}$ & $\Delta t=\frac{1}{400}$  \\ \hline
Error & 6.62e-3 & 1.98e-3 & 1.34e-3 & 5.94e-3 & 2.08e-3
    \\ \hline
   95\% C.I. & 7.22e-3 & 9.54e-3 & 9.48e-3 & 5.52e-3 & 5.07e-3 
    \\ \hline
M.S.E. & 5.74e-5 & 2.76e-5 & 2.55e-5 & 4.26e-5 &7.77e-6  \\ \hline
\end{tabular}
\caption{Error in numerical empirical averages with 95 \% confidence intervals and mean squared errors. The reference value is $0.640172$. }
  \label{table:test2_1}
\end{table}

We also carry out the $\Delta t$ test for the test potential
\eqref{ex:pot1} with an off-diagonal observable:
\begin{equation} \label{eq:ob2}
\wh{A}=\begin{bmatrix}
0   &    e^{- {\wh q}^2}\\
  e^{- {\wh q}^2} &   0  
\end{bmatrix}
\end{equation}
We take $\beta=1$, $M=10$ and $\eta=5$, and test time step
sizes $\Delta t = \frac{1}{25}$, $\frac{1}{50}$, $\frac{1}{100}$,
$\frac{1}{200}$ and $\frac{1}{400}$ till $T=10000$. The numerical
results are plotted in Figure \ref{fig:diffdt}. The errors in the empirical averages
together with their $95\%$ confidence intervals and mean squared errors are shown in Table
\ref{table:test2_2}. In this test, when $\Delta t$ is large, the
numerical error seems to be dominated by error in numerical
integration of the trajectory, and reducing $\Delta t$ reduces the
error of the sampling. Compared with the previous case with diagonal
observables, the variance of the observable is larger even when a
small time step size is used; sampling the thermal average for
observables with off-diagonal entries is more challenging. This is
easy to understand as for off-diagonal observables, the ring polymer
has to have a kink to make the weight function non-zero, and hence the
surface hopping dynamics becomes more important. As we will see in the
next subsection, increasing the hopping parameter $\eta$ in the
PIMD-SH can improve the results.
\begin{figure}[ht]
\begin{center}
\includegraphics[scale=0.6]{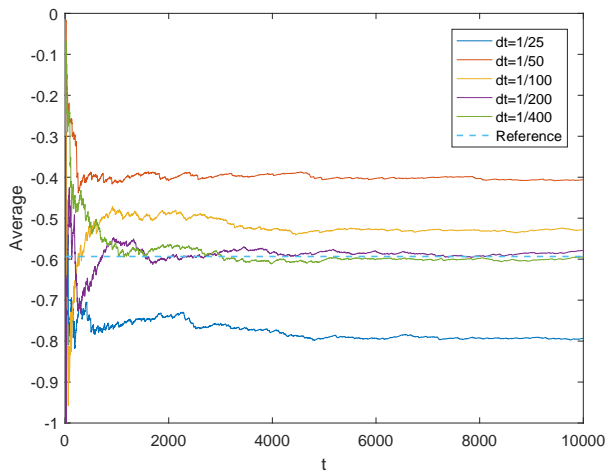}
\end{center}
\caption{Running average of an off-diagonal observable with test
  potential~\ref{ex:pot1} for different time step size $\Delta t$.  }
\label{fig:diffdt2}
\end{figure}
\begin{table}
  \centering
  \begin{tabular}{ c| c| c|c|c|c} \hline
    & $\Delta t=\frac{1}{25}$ & $\Delta t=\frac{1}{50}$ & $\Delta t=\frac{1}{100}$ & $\Delta t=\frac{1}{200}$ & $\Delta t=\frac{1}{400}$  \\ \hline
Error & 1.99e-1 & 1.87e-1 &  6.54e-2 & 1.50e-2 & 1.85e-3
    \\ \hline
   95\% C.I. & 2.70e-2 & 2.09e-2 & 2.46e-2 & 2.05e-2 &  1.14e-2 
    \\ \hline
M.S.E. & 3.99e-2 & 3.51e-2 & 4.43e-3 & 3.34e-4 & 3.74e-5  \\ \hline
\end{tabular}
\caption{Errors in numerical empirical averages with 95 \% confidence intervals and mean squared errors. The reference value is $-0.593497$. }
  \label{table:test2_2}
\end{table}

\subsection{Effect of the hopping intensity parameter $\eta$} \label{mutest}
In this test, we implement the PIMD-SH method with the test potential \eqref{ex:pot1}, with $\beta=1$,  $\beta_N=\frac{1}{16}$ and $M=10$. We choose the off-diagonal  observable \eqref{eq:ob2} as before.

As we mentioned before in \S\ref{sec:testex} and also have seen in the
numerical results in Figures~\ref{fig:diffdt} and \ref{fig:diffdt2},
sampling of the off-diagonal observable is more challenging in PIMD-SH
due to the contribution to the weight by ring polymer configurations
with kinks and also the sampling variance is larger. In this test, we
show that increasing the hopping intensity parameter $\eta$, which
makes hopping more frequent, helps sampling off-diagonal observables.
For $\eta=5$, $10$, $20$ and $40$, we test the PIMD-SH method with
$\Delta t= (100\eta)^{-1}$ respectively till $T=10000$. Recall that a
small time step size is required to maintain the accuracy and
stability of the integrator.

\begin{figure}[ht]
\begin{center}
\includegraphics[scale=0.6]{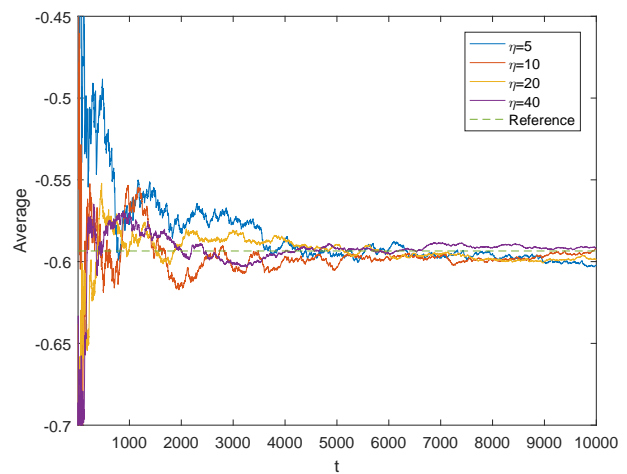}
\end{center}
\caption{Running average of the off-diagonal observable with different $\eta$. }
\label{fig:diffeta}
\end{figure}

The results are plotted in Figure \ref{fig:diffeta} and the errors in
the empirical averages together with their $95 \%$ confidence
intervals and mean squared errors are shown in Table \ref{table:3}.
Increasing the hopping intensity parameter can effectively reduce the
sampling error and variance. We admit though it is computationally
more expensive to use a larger $\eta$ since the time step size has to
be smaller by directly applying the JBAOABJ scheme. A better numerical
scheme in the spirit of \cite{LuVandenEijnden:13, YuLuAbramsVE:16} is
needed and will be leaved for future works.
\begin{table} \label{table:3}
  \centering
  \begin{tabular}{ c| c| c|c|c} \hline
    & $\eta=5$ & $\eta=10$ & $\eta=20$ & $\eta=40$   \\ \hline
 Error & 8.27e-3 & 1.38e-3 & 4.97e-3 & 1.44e-3
    \\ \hline
   95\% C.I. & 1.09e-2 & 7.67e-3 & 3.59e-3 & 1.87e-3
    \\ \hline
M.S.E. & 9.93e-5 & 1.32e-5 & 2.80e-5 & 2.98e-6   \\ \hline
\end{tabular}
\caption{Errors in numerical empirical averages with 95 \% confidence intervals and mean squared errors. The reference value is $-0.593497$. }
  \label{table:test3_1}
\end{table}

\section{Conclusion}

We have proposed in this work the PIMD-SH method for sampling the thermal equilibrium average of multi-level quantum systems. The formulation is justified theoretically and supported by numerical results. 

Among the possible future directions based on the current work, the most interesting direction is perhaps to combine this approach with surface hopping method for sampling dynamical correlation functions of the type $\tr_{ne}[e^{-\beta \wh{H}} \wh{A}(t) \wh{B}]$, for which the ring polymer representation we developed in this work is well suited.

Better numerical integration strategies especially for using a larger
hopping intensity parameter $\eta$ is worth exploring.  Further
investigation is also needed when the off-diagonal components of the
potential function changes sign or takes complex values.

\begin{acknowledgments}
  This work is partially supported by the National Science Foundation
  under grant DMS-1454939. J.L. would like to thank Giovanni Ciccotti
  and John Tully for helpful discussions.
\end{acknowledgments}

\appendix

\section{Ring polymer representations for a general two-level matrix potential} \label{App:A}

We have assumed that the off-diagonal terms of the matrix potential in a  two-level system $V_{01} = V_{10}$ is real and does not change sign. In this Appendix, we discuss the formulations for a general two-level system. Since the potential matrix is Hermitian, the diagonal potential terms $V_{00}$ and $V_{11}$ are always real, while $V_{01}$ and $V_{10}$ may be complex with $V_{10}=V_{01}^{\ast}$. 

If we repeat the derivation in Section \ref{sec:derivation} for the
general case, everything is parallel till the step when we approximate
$e^{-\beta_N V}$ with the Strang splitting in \eqref{eq:expV}. In the general case, the off
diagonal part of the matrix changes to
\[
V_o=
\begin{bmatrix}
  0   &    V_{01} \\
  \bar{V}_{01} & 0
\end{bmatrix}.
\]
Using again the Strang splitting and the explicit expression of
$\exp(-\beta_N V_o)$, we get (cf. \eqref{eq:entryexpV})
\begin{multline}
  \bra{\ell} e^{-\beta_N V} \ket{\ell'}
  = \\ = \begin{cases}
    e^{-\beta_N V_{\ell \ell}+ \ln \bigl( \cosh(\beta_N \abs{V_{01}}) \bigr)} + \Or(\beta_N^3), 
    & \ell = \ell'; \\
    - \frac{V_{\ell\ell'}}{\abs{V_{\ell\ell'}}} e^{-\beta_N\frac{V_{00}+V_{11}}{2} + \ln\bigl( \sinh(\beta_N \abs{V_{01}}) \bigr) } + \Or(\beta_N^3), & \ell \neq \ell'. 
    \end{cases}
\end{multline}
Different from the case considered in the main text, since
$V_{\ell\ell'}$ is complex, in general the phase factor does not
cancel for all the kinks in the ring polymer. We can still approximate
the partition function as an average over the extended phase space,
though the quantities to be averages is now complex (in the case where
the off-diagonal entry of $V$ is real but changes sign, we will need
to average over terms with different signs). If we collect of the
phase factors in a single term, we arrive at the following
approximation 
\begin{multline*}
  \tr_{ne}[e^{-\beta \wh H} ] = \frac{1}{(2\pi)^{dN}} \int_{\RR^{2dN}}
  \ud \bd q \ud \bd p \, \sum_{\bd{\ell}\in \{ 0,1\}^N} \\ \times
  \exp\left( -\beta_N \sum_{k=1}^N \langle \ell_k | G_{k}| \ell_{k+1}
    \rangle \right) W_N[I] +\Or(N \beta_N^3).
\end{multline*}
where the weight factor becomes
\[
W_N[I]= \prod_{k=1}^N \langle \ell_k | J_{k}| \ell_{k+1} \rangle, 
\]
and $G_k$, $J_k$ are still defined as before in \eqref{eq:Gele} and
\eqref{eq:Jele} respectively. Recall that in the previous case, the
weight factor for the partition function, corresponding to the
identity operator, is just constant $1$. As a result, the thermal
average of a general observable becomes a ratio of ensemble averages
\[
\langle \wh{A} \rangle  \approx  \dfrac{\dps\int_{\RR^{2dN}} \ud \bd q  \ud \bd p  \sum_{\bd{\ell}\in\{0, 1\}^N} \pi (\widetilde {\bd{z}} )W_N[A] (\widetilde {\bd{z}})}{\dps\int_{\RR^{2dN}} \ud \bd q  \ud \bd p  \sum_{\bd{\ell}\in\{0, 1\}^N} \pi (\widetilde {\bd{z}} )W_N[I] (\widetilde {\bd{z}})},
\]
where the expression of $W_N[A]$ is same as before in \eqref{eq:reA}
and the equilibrium distribution is given in \eqref{eq:pi}. This can
still be sampled using a PIMD-SH dynamics associated with the Gibbs
distribution $\pi$ on $\RR^{2dN} \times \{0, 1\}^N$, though some bias
will be introduced if we use the same ergodic trajectory to sample
both the numerator and denominator. 

Another consequence of the general phase of the term
$-V_{01}/ \abs{V_{01}}$ is that the partition function is determined
through averaging of terms that change signs / phases, which is
reminiscent of the ``sign problems'' in quantum Monte Carlo
simulations. Further investigations of such issues are needed.

Let us remark that there are other possibilities of choosing
the effective Hamiltonian and the corresponding weight functions. For
example, after the splitting of $e^{-\beta_N V}$, instead rewrite the
$\sinh$ and $\cosh$ functions to the exponent, we can put these terms
to the weight functions. We choose the present approach since in the
case that the off-diagonal matrix potentials are real and do not
change sign, it reduces to a nice probabilistic sampling problem; but
for general cases, it is worth considering other approaches.  

\section{Ring polymer representations for $M$-level system
  ($M\ge3$)} \label{App:B} 

We show the extension of the ring polymer representation to a general
$M$-level system ($M\ge2$) in this Appendix. For simplicity of the
presentation, let us just focus on the expression for the partition
function, cf.~the derivation in Section \ref{sec:derivation}; the
expression for the observable follows analogously. For the $M$-level
case, The Hamiltonian is given by
\[
\wh H 
= \wh T + \wh V ,
\]
where the kinetic operator is still diagonal, and the potential matrix is a $M\times M$ Hermitian matrix such that $V_{ij}=V_{ji}^*,\,i,j=1,\cdots,M$. We decompose the matrix potential potential as follows
\[
V =V_d+ \sum_{j<i =1}^M V^{(i,j)},
\]
where $V_d$ is the diagonal part of $V$ and the off-diagonal part has been decomposed pairwisely into a sum of $V^{(i,j)}$  with elements given by
\begin{multline}
  \bra{\ell} V^{(i,j)} \ket{\ell'} = \\ = \begin{cases}
    V_{\ell \ell'}, \quad \ell=i,\,\ell'=j,\ \text{or} \ \ell=j,\,\ell'=i;  \\
    0, \quad \quad \text{otherwise}.
    \end{cases}
\end{multline}
Note that  in reformulating the partition function, one can still apply the Strang splitting to each $e^{-\beta_N \wh H}$ and obtain \eqref{eq:inter1}. Next, in order to derive an approximate formula for matrix elements of $e^{-\beta_N V}$, we applied the Strang splitting multiple times and get
\begin{equation}\label{eq:expVm}
\begin{aligned}
  e^{-\beta_N V} & = e^{-\beta_N \big(V_d +  \sum_{j<i} V^{(i,j)}\big)} \\
                 & =  e^{-\beta_N V^{(M,M-1)}/2 }\cdots e^{-\beta_N V^{(2,1)} /2}e^{-\beta_N V_d } \cdots \\
& \quad  \times e^{-\beta_N V^{(2,1)} /2}\cdots  e^{-\beta_N V^{(M,M-1)}/2 }+ \mathcal O (\beta_N^3)\\
&=: M_v+ \mathcal O (\beta_N^3),
\end{aligned}
\end{equation}
where the last line defines $M_v$, the approximate matrix for $e^{-\beta_N V}$.
Here, for the diagonal part, 
\begin{equation}
  \bra{\ell} e^{-\beta_N V_d } \ket{\ell'}
 = \begin{cases}
e^{ -\beta_N V_{\ell \ell'} }, \quad \ell=\ell'; \\
0, \quad \quad \ell \ne \ell'.
    \end{cases}
\end{equation}
And for the off-diagonal part, we have
\begin{multline}
  \bra{\ell} e^{-\beta_N V^{(i,j)} } \ket{\ell'}
   = \\ =  \begin{cases}
\cosh(\beta_N |V_{i,j}|), \quad\quad\quad \quad \ell=\ell' \in \{ i,j\}; \\
-\frac{V_{\ell \ell'}}{|V_{\ell \ell'}|}\sinh(\beta_N |V_{\ell \ell'}|), \quad \ell=i, \ell'=j\, \text{or}\,\ell=j, \ell'=i; \\
1,\quad\quad\quad\quad\quad\quad\quad\quad\quad \,\, \ell=\ell' \notin  \{ i,j\};\\
0, \quad \quad\quad\quad\quad \quad\quad \quad\quad \, \, \text{otherwise}.
    \end{cases}
\end{multline}
Next, we observe that
\begin{equation}\label{eq:expVexp}
\begin{aligned}
  \bra{\ell} M_v \ket{\ell'} & =  \bra{\ell} e^{-\beta_N V^{(M,M-1)}/2 }\cdots e^{-\beta_N V^{(2,1)}/2 }e^{-\beta_N V_d } \cdots \\
& \quad \quad  e^{-\beta_N V^{(2,1)} /2}\cdots  e^{-\beta_N V^{(M,M-1)}/2 } \ket{\ell'} \\
& = \sum_{\bd n}   \bra{\ell} e^{-\beta_N V^{(M,M-1)} /2}\ket{n_{(M,M-1)}}\times \cdots \\
& \quad \cdots \times \bra{n_{(2,2)}}e^{-\beta_N V^{(2,1)} /2}\ket{n_{(2,1)}} \times \\
& \quad \times \bra{n_{(2,1)}} e^{-\beta_N V_d } \ket{n_{(2,1)}'} \times \\
& \quad \times \bra{n_{(2,1)}'}e^{-\beta_N V^{(2,1)} /2}\ket{n_{(2,2)}'} \times \cdots\\
& \quad \cdots  \bra{n_{(M,M-1)}'} e^{-\beta_N V^{(M,M-1)} /2}\ket{\ell},
\end{aligned}
\end{equation}
where $\bd n = (n_{(M,M-1)},\cdots,n_{(2,1)},n_{(2,1)}',\cdots,n_{(M,M-1)}' )$ with each entry takes possible values in $\{1, \ldots, M\}$. We introduce the augmented index $\overline {\bd n} = \{ \ell, \bd n, \ell' \} $, and observe that when two consecutive index in $\overline {\bd n} $ are different, the product in \eqref{eq:expVexp} is either $0$ or gains a multiplier of order $\mathcal O(\beta_N)$. Therefore, if we omit all the terms which are of order $\mathcal O (\beta_N^3)$, we conclude that, when $\ell'=\ell$,
\begin{equation}\label{eq:expVm1}
\begin{aligned}
 \bra{\ell} e^{-\beta_N V} \ket{\ell} & =\bra{\ell} M_v \ket{\ell}+ \mathcal O (\beta_N^3) \\
& =e^{-\beta_N V_{\ell,\ell}}\sum_{\bar \ell\in\{0,\cdots,M\}} f_{\bar \ell,\ell}+ \mathcal O (\beta_N^3),
\end{aligned}
\end{equation}
where
\begin{equation*}
 f_{\bar \ell,\ell}
 = \begin{cases}
\cosh\left( \frac {\beta_N\sqrt 2}{2} \sqrt{\sum_{\ell' \ne \ell} |V_{\ell, \ell'}|^2} \right), \quad \bar\ell=\ell; \\
2 \sinh\left( \frac{\beta_N^2}{4}|V_{\ell,\bar \ell}|^2 \right), \quad \quad \bar \ell \ne \ell.
    \end{cases}
\end{equation*}
By the Taylor's expansion, we can further simplify and obtain
\begin{equation}
 \bra{\ell} e^{-\beta_N V} \ket{\ell} =e^{-\beta_N V_{\ell,\ell}} \cosh\Big( \beta_N \sqrt{\sum_{\ell' \ne \ell} |V_{\ell, \ell'}|^2} \Big)+ \mathcal O (\beta_N^3).
\end{equation}
Similarly, for $\ell'\neq\ell$, if we omit all the terms which are of order $\mathcal O (\beta_N^3)$ in \eqref{eq:expVexp}, we obtain that
\begin{equation}
\begin{aligned}
 \bra{\ell} e^{-\beta_N V} \ket{\ell'}& =-\frac{V_{\ell,\ell'}}{|V_{\ell,\ell'}|} \sinh\left(\frac{\beta_N}{2} |V_{\ell,\ell'}|\right) \\
& \qquad \times \left(e^{-\beta_N V_{\ell,\ell}}+e^{-\beta_N V_{\ell',\ell'}} \right)+ \mathcal O (\beta_N^3) \\
&=-\frac{V_{\ell\ell'}}{|V_{\ell\ell'}|} \sinh\left(\beta_N |V_{\ell,\ell'}|\right)e^{- \beta_N \frac{V_{\ell,\ell}+ V_{\ell',\ell'}}{2}} \\
  & \qquad + \mathcal O (\beta_N^3)
\end{aligned}
\end{equation}
Note that the two expressions above are natural generalization of 
the two level counterpart \eqref{eq:entryexpV}.

\section{Ring polymer representations for $2$-level system in the adiabatic picture} \label{App:C}
In this section, we instead use the adiabatic basis to handle discrete electronic states, and we are able to recover the results by Schmidt and Tully in \cite{SchmidtTully2007}. We start with \eqref{eq:inter1}, where the electronic states have not been specified yet.
We denote the adiabatic states by $\Phi_n(q)$ and the adiabatic surface by $E_n(q)$, where $n=0,1$, and they satisfy
\begin{equation}
V(q) \Phi_n(q)=E_n(q)\Phi_n(q). 
\end{equation}
By inserting multiple times the resolution of identity $I=\sum_{n=0,1}\ket{\Phi_n (q)}\bra{\Phi_n (q)}$, we get
\begin{align*}
 \tr_{ne}[e^{-\beta \wh H}]  
&= \frac{1}{(2\pi)^{dN}} \int_{\RR^{2dN}} \ud \bd q  \ud \bd p \\
&\quad \times  \sum_{\bd\ell\in \{0,1\}^N} \bra{\Phi_{\ell_1} (q_1)} e^{-\beta_N  V(q_1)  } \ket{\Phi_{\ell_2} (q_2)}\times \\
&\quad \times  \bra{\Phi_{\ell_2} (q_2)} e^{-\beta_N V(q_2)} 
\ket{\Phi_{\ell_3} (q_3)} \cdots \\
&\quad \times   \bra{\Phi_{\ell_N} (q_N)} e^{-\beta_N  V(q_N)  } \ket{\Phi_{\ell_1} (q)}\times  \\
& \quad \times e^{-\beta_N \bigl(S_1+ \cdots S_N \bigr)}+\mathcal O(N \beta_N^3)\\
\end{align*}
As $\Phi_n$ are eigenfunctions of $V$, $e^{-\beta_N V(q)} \Phi_n(q) =e^{-\beta_N E_n(q)}\Phi_n(q)$,
the above can be further simplified as 
\begin{align*}
& \quad \tr_{ne}[e^{-\beta \wh H}]  \\
&=  \frac{1}{(2\pi)^{dN}} \int_{\RR^{2dN}} \ud \bd q  \ud \bd p  \sum_{\bd\ell\in \{0,1\}^N} \prod_{j=1}^N   \langle{\Phi_{\ell_j} (q_j)|\Phi_{\ell_{j+1}} (q_{j+1})}\rangle \times \\
& \quad \times  e^{-\beta_N \bigl(E_{n_1}+\cdots E_{n_{j+1}}+S_1+ \cdots S_N \bigr)}+\mathcal O(N \beta_N^3).
\end{align*}
If we define the short hand
\[
\psi (\bd q, \bd \ell) =  \prod_{j=1}^N   \langle{\Phi_{\ell_j} (q_j)|\Phi_{\ell_{j+1}} (q_{j+1})}\rangle,
\] 
we obtain 
\begin{multline}
\tr_{ne}[e^{-\beta \wh H}] = \frac{1}{(2\pi)^{dN}} \int_{\RR^{2dN}} \ud \bd q  \ud \bd p  \sum_{\bd\ell\in \{0,1\}^N} \\
\times \frac{\psi (\bd q, \bd \ell)}{|\psi (\bd q, \bd\ell)|} e^{-\beta_N \overline{H}_N(\bd q, \bd p, \bd \ell) },
\end{multline}
with the effective Hamiltonian given by 
\begin{multline*}
\overline{H}_N(\bd{q}, \bd{p}, \bd{\ell}) = \sum_{k=1}^N \left( \frac{p^2_k}{2M}+ \frac{M }{2 \beta_N^2} \left(q_k-q_{k+1} \right)^2 + E_{\ell_k} \right) \\
- \frac{1}{\beta_N} \ln |\psi (\bd q, \bd \ell)|.
\end{multline*}
This is exactly the ring polymer representation derived in
\cite{SchmidtTully2007}.  The PIMD-SH sampling method can be applied
to the adiabatic picture as well, which we will consider in future
works. 

\bibliography{surfacehoppingPIMD} 
\end{document}